\documentclass[prd,twocolumn,superscriptaddress,altaffilletter,showpacs]{revtex4}
\usepackage{amssymb,amsmath}
\usepackage{comment}
\usepackage{graphicx}
\DeclareUnicodeCharacter{2212}{\textendash}

\begin{document}

\title{Kerr black hole in de Sitter spacetime and observational redshift: \\
Toward a new method to measure the Hubble constant}
\author{Mehrab Momennia}
\email{mmomennia@ifuap.buap.mx, momennia1988@gmail.com}
\affiliation{Instituto de F\'{\i}sica, Benem\'erita Universidad Aut\'onoma de Puebla,\\
Apartado Postal J-48, 72570, Puebla, Puebla, Mexico}
\author{Alfredo Herrera-Aguilar}
\email{aherrera@ifuap.buap.mx}
\affiliation{Instituto de F\'{\i}sica, Benem\'erita Universidad Aut\'onoma de Puebla,\\
Apartado Postal J-48, 72570, Puebla, Puebla, Mexico}
\author{Ulises Nucamendi}
\email{ulises.nucamendi@umich.mx, unucamendi@gmail.com}
\affiliation{Instituto de F\'{\i}sica y Matem\'{a}ticas, Universidad Michoacana de San
Nicol\'as de Hidalgo,\\
Edificio C--3, Ciudad Universitaria, CP 58040, Morelia, Michoac\'{a}n, Mexico}
\date{\today }

\begin{abstract}
We extract the Hubble law by the frequency-shift considerations of test
particles revolving the Kerr black hole in asymptotically de Sitter
spacetime. To this end, we take into account massive geodesic particles
circularly orbiting the Kerr-de Sitter black holes that emit redshifted
photons towards a distant observer which is moving away from the
emitter--black hole system. By considering this configuration, we obtain an
expression for redshift in terms of the spacetime parameters, such as mass,
angular momentum, and the cosmological constant. Then, we find the frequency
shift of photons versus the Hubble constant with the help of some physically
motivated approximations. Finally, some exact formulas for the Schwarzschild
black hole mass and the Hubble constant in terms of the observational
redshift of massive bodies circularly orbiting this black hole are
extracted. Our results suggest a new independent general relativistic
approach to obtaining the late-time Hubble constant in terms of observable
quantities. \vskip3mm

\end{abstract}

\maketitle

%: circular equatorial orbits}

%%%%%%%%%%%%%%%%%%%%%%%%%%%%%%%%%%%%%%%%%%%%%%%%%%%%%%%%%%%%%%%%%%%%

\section{Introduction}

Black holes are the densest massive objects known in nature and are among
the most important and interesting solutions to the Einstein field
equations. By now, they have been directly detected through gravitational
waves produced by the coalescence events captured in the LIGO and Virgo
observatories \cite{GW} as well as the shadow images of supermassive black
holes hosted at the center of the Milky Way galaxy and the M87 galaxy
revealed by the EHT Collaboration \cite{EHTM87,EHTSgr}. Therefore, nowadays,
exploring various aspects of black holes' physics attracts much attention in
the context of the general relativity theory.

Among others, inventing and developing methods to determine the black hole
parameters, such as mass, charge, and angular momentum has a special place.
One of the robust methods to obtain the black hole parameters was initially
suggested in \cite{PRDhn}, and then developed to analytically express the
mass and spin parameters of the Kerr black hole in terms of a few directly
observable quantities \cite{PRDbhmn}. In this general relativistic
formalism, the observables are frequency shifts of photons emitted by
massive geodesic particles orbiting the central black holes along with their
orbital parameters.

From the theoretical point of view, the method of \cite{PRDhn}\ has been
applied to several black hole spacetimes, such as higher--dimensional
Myers--Perry black holes \cite{MyersPerry}, Kerr--Newman black holes in de
Sitter (dS) spacetime \cite{KNdS}, the Plebanski--Demianski background \cite%
{PlebanskiDemianski}, and spherically symmetric regular black holes \cite%
{RegularBH}.\ In addition, the boson stars \cite{BosonStar}\ as well as
black holes in modified gravity \cite{MOG}, coupled to nonlinear
electrodynamics \cite{NED}, and immersed in a strong magnetic field \cite%
{SMF} have been investigated by employing a similar procedure, i.e. finding
a relation between frequency shift and compact object parameters. However,
all the aforementioned attempts were based on the kinematic redshift which
is not a directly measured observational quantity, unlike the total
frequency shift of photons. Thus, this fact has motivated us to take into
account the total redshifts of photons and obtain concise and elegant
analytic formulas for the mass and spin of the Kerr black hole in terms of
these directly observable elements \cite{PRDbhmn}. More recently, this
method was also applied to express the parameters of static polymerized
black holes in terms of the total frequency shifts \cite{FuZhang}.

From a practical point of view, the developed prescription of this general
relativistic approach has been employed to estimate the mass-to-distance
ratio of some supermassive black holes hosted at the core of active galactic
nuclei (AGNs), like NGC 4258 \cite{ApJL}, TXS-2226-184 \cite{TXS}, and an
additional $15$ galaxies \cite{TenAGNs,FiveAGNs}. These AGNs enjoy accretion
disks consisting of water vapor clouds that are circularly orbiting the
central supermassive black hole and emitting photons toward the distant
observer, hence enabling us to estimate the mass-to-distance ratio and
quantify the gravitational redshift produced by the spacetime curvature that
is a general relativistic effect.

On the other hand, the so-called $\Lambda $-cold dark matter cosmological
standard model successfully explains the current epoch in the evolution of
the cosmos. The field equations of Einstein gravity in the presence of a
cosmological constant $\Lambda $ along with an energy-momentum tensor $%
T^m_{\mu \nu } $ that accounts for the matter content of the Universe read%
\begin{equation}
G_{\mu \nu }+\Lambda g_{\mu \nu }=T^m_{\mu \nu } ,  \label{FEs}
\end{equation}%
where $G_{\mu \nu }$\ is the Einstein tensor. Thus, in order to explain the
current accelerated expansion of the Universe, taking into account the
contribution of the dark energy, and thus adding the $\Lambda $-term to the
Einstein field equations is inevitable \cite{DarkEnergy,Carroll}. Indeed,
although the observations in small scales could be explained by the first
term of the left-hand side (lhs) of Eq. (\ref{FEs}), a consistent
description of the large-scale structure of the Universe requires
considering the second term. Therefore, it is quite natural to attempt to
quantitatively clarify the influence of the repulsive cosmological constant
on the detected redshift and blueshift of photons coming from massive
geodesic particles, stars for instance, orbiting the Kerr black hole.

In order to advance work in this direction, we shall consider the field
equations (\ref{FEs}) in the absence of matter content as the first step. A
family of solutions to these simplified field equations describes black
holes in asymptotically dS spacetime. %\cite{Carter}. 
Moreover, the rotating black hole solutions to the Einstein-$\Lambda $ field
equations are described by the Kerr-dS (KdS) line element \cite{KdS} and the
properties of the geodesic motion in this background have been investigated
in \cite{KdSECO,Hackmann,Kraniotis} (see \cite{KNdS} as well). Thus, by
taking into account the Universe expansion effect encoded in the
cosmological constant through the explicit appearance of the $\Lambda $ term
in the metric, we push forward the formalism developed in \cite%
{PRDhn,PRDbhmn} for expressing the Kerr black hole parameters in terms of
purely observational quantities to the case in which the Hubble constant can
also be determined.

The consideration of the accelerated expansion of the Universe in the
redshift due to a cosmological constant has potential interest in terms of
astrophysical applications, since many of the AGNs with megamaser disks
orbiting its central black hole are within the Hubble flow \cite%
{MCPI,MCPIII,MCPV,MCPVI,MCPVIII,MCPIX,MCPXI,MCPXII,MCPXIII}. Therefore, this
modeling includes the contribution of the expansion of the Universe in the
metric, making it suitable for describing this effect on the total redshift
of photons emitted by test particles and detected on Earth. Thus, this new
form of accounting for the dS accelerated expansion of the Universe in the
expression for total redshift allows us to extract the Hubble law as well.
Finally, it is worth noticing that this approach differs from the previous
ones in which the expansion effect is taken into account in the total
redshift through a composition of redshifts that has no metric origin (see,
for instance, \cite{MCPXI,MCPXIII,TenAGNs,FiveAGNs}).

The outline of this paper is as follows. The next section is devoted to a
brief review of the geometrical properties of the KdS black holes and the
geodesic motion in this background. Besides, we analytically obtain the
valid parameter space for having KdS black holes, and also review our
general relativistic formalism that allows expressing the black hole
parameters in terms of observational redshift. In Sec. \ref%
{frequencyVSparameters}, we express the redshift of emitters that are
circularly orbiting the KdS black holes in terms of the parameters of
spacetime while the detector is in radial motion. Then, by considering a
physically motivated configuration, we extract the Hubble law in its
original formulation from the obtained frequency shift relation. Finally, we
find analytic expressions for the Schwarzschild black hole mass and the
Hubble constant in terms of the observational frequency shifts of photons
emitted by massive particles orbiting circularly a Schwarzschild black hole.
We finish our paper with some concluding remarks.

\section{Kerr-de Sitter spacetime}

Here, we give a short review of the geometrical properties of the rotating
black holes in the dS background and analytically obtain valid parameter
space for having KdS black holes in Sec. \ref{KerrDS}. Then, we study the
geodesic motion of massless/massive particles in this geometry in Sec. \ref%
{Geodesics} and derive equations\ that are important for our next purposes.
Finally, in Sec. \ref{FrequencyShift}, we briefly review our general
relativistic formalism that allows us to express the black hole parameters
in terms of observational redshift and orbital parameters of massive
geodesic particles orbiting around the black holes. We shall use the general
results of this section for a special configuration in Sec. \ref%
{frequencyVSparameters} to extract the Hubble law.

\subsection{Properties of the Kerr-dS background}

\label{KerrDS}

The KdS line element in the standard Boyer-Lindquist coordinates $\left(
t,r,\theta ,\varphi \right) $ reads \cite{KdS} (we use $c=1=G$ units) 
\begin{equation}
ds^{2}=g_{tt}dt^{2}+2g_{t\varphi }dtd\varphi +g_{\varphi \varphi }d\varphi
^{2}+g_{rr}dr^{2}+g_{\theta \theta }d\theta ^{2},  \label{metric}
\end{equation}%
with the metric components%
\begin{equation}
g_{tt}=-\left( \frac{\Delta _{r}-\Delta _{\theta }a^{2}\sin ^{2}\theta }{%
\Sigma \Xi ^{2}}\right) ,\quad g_{rr}=\frac{\Sigma }{\Delta _{r}},\quad
g_{\theta \theta }=\frac{\Sigma }{\Delta _{\theta }},  \label{grr}
\end{equation}%
\begin{equation}
g_{\varphi \varphi }=\frac{\sin ^{2}\theta }{\Sigma \Xi ^{2}}\left[ \Delta
_{\theta }\left( r^{2}+a^{2}\right) ^{2}-\Delta _{r}a^{2}\sin ^{2}\theta %
\right] \,,
\end{equation}%
\begin{equation}
g_{t\varphi }=-\frac{a\sin ^{2}\theta }{\Sigma \Xi ^{2}}\left[ \Delta
_{\theta }\left( r^{2}+a^{2}\right) -\Delta _{r}\right] ,
\end{equation}%
where the functions $\Delta _{r}\left( r\right) $, $\Delta _{\theta }\left(
\theta \right) $, $\Sigma \left( r,\theta \right) $, and $\Xi $ have the
following explicit form 
\begin{equation}
\Delta _{r}=r^{2}+a^{2}-2Mr-\frac{\Lambda r^{2}}{3}\,\left(
r^{2}+a^{2}\right) ,  \label{DeltaR}
\end{equation}%
\begin{equation}
\Delta _{\theta }=1+\frac{\Lambda }{3}a^{2}\cos ^{2}\theta ,
\end{equation}%
\begin{equation}
\Sigma =r^{2}+a^{2}\cos ^{2}\theta \,,
\end{equation}%
\begin{equation}
\Xi =1+\frac{\Lambda }{3}a^{2},
\end{equation}%
and $M$ is the total mass of the black hole, $a$ is the angular momentum per
unit mass $a=J/M$, and $\Lambda $\ is the cosmological constant related to
dS radius $l_{dS}$ as $\Lambda =3/l_{dS}^{2}$. The KdS metric (\ref{metric})
describes an axially symmetric and stationary spacetime (between the event
horizon and the cosmological horizon)\ that reduces to the standard Kerr
black hole in the limit $\Lambda =0$ and the Schwarzschild-dS (SdS) black
hole for $a=0$. The coordinate singularities of this spacetime are
characterized by $\Delta _{r}=0$ (the four roots correspond to horizons),
while calculation of the invariant curvature scalar reveals that the
intrinsic singularity is given by $\Sigma =0$ where located at $\left\{
r=0,\theta =\pi /2\right\} $ for $a\neq 0$. Therefore, the presence of a
cosmological horizon, characterized by the largest root of $\Delta _{r}$, is
one of the consequences of non-vanishing $\Lambda $.

In order to find the extreme values of $a$\ and $\Lambda $\ for having black
holes, we found that it is convenient to introduce the normalized variables $%
x$, $\alpha $, and $\lambda $ as below%
\begin{equation}
x=\frac{r}{M},\quad \alpha =\frac{a}{M},\quad \lambda =\frac{\Lambda M^{2}}{3%
},
\end{equation}%
and express the $\Delta _{r}$\ function in terms of the new variables as
follows%
\begin{equation}
M^{-2}\Delta _{r}=x^{2}+\alpha ^{2}-2x-\lambda x^{2}\,\left( x^{2}+\alpha
^{2}\right) .  \label{DeltaFun}
\end{equation}

%%%%%%%%%%%%%%%%%%%%%%% Horizons %%%%%%%%%%%%%%%%%%%%%%%%%%%%%%%%%%%%%%%%
\begin{figure}[tbp]
\centering
\includegraphics[width=0.45\textwidth]{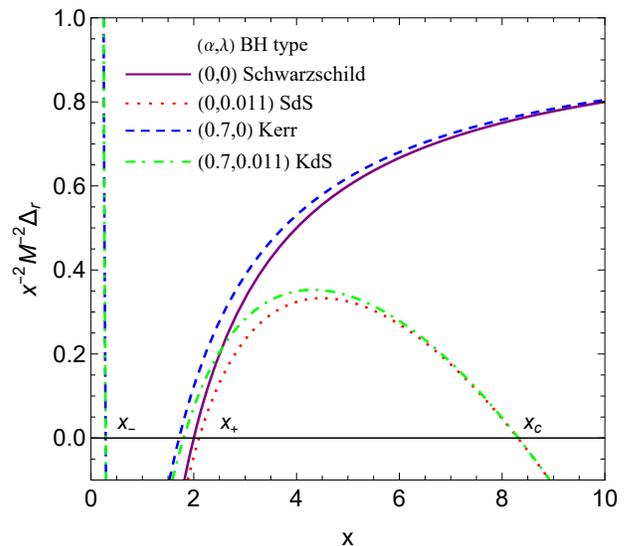}
\caption{The general behavior of $\Delta _{r}$ function given in (\protect
\ref{DeltaFun}) versus the radial coordinate $x$ for four black hole types,
namely Schwarzschild, SdS, Kerr, and KdS. $\Delta _{r}$ is positive between $%
x _{+}$ and $x _{c}$ as well as before $x _{-}$, whereas is negative
otherwise. By increasing $\protect\lambda$ ($\protect\alpha$), $x _{c}$ ($x
_{-}$) approaches $x _{+}$ (not shown here).}
\label{DelFunFig}
\end{figure}
%%%%%%%%%%%%%%%%%%%%%%%%%%%%%%%%%%%%%%%%%%%%%%%%%%%%%%%%%%%%%%%

Generally, one can show that $\Delta _{r}$ has (at most) four distinct roots
that can be regarded as the cosmological horizon ($x=x_{c}$), the event
horizon ($x=x_{+}$), the inner horizon ($x=x_{-}$), and a negative root ($%
x=x_{0}<0$) so that $x_{-}<x_{+}<x_{c}$ (see Fig. \ref{DelFunFig}).
Therefore, we can express $\Delta _{r}$ in terms of these quantities in the
following form%
\begin{equation}
M^{-2}\Delta _{r}=\lambda \left( x-x_{0}\right) \left( x-x_{-}\right) \left(
x-x_{+}\right) \left( x_{c}-x\right) .  \label{DeltaZeros}
\end{equation}

Now, by equating the equations for $\Delta _{r}$ given in (\ref{DeltaFun})\
and (\ref{DeltaZeros}), the following relations between parameters are found%
\begin{equation}
\alpha ^{2}=\lambda x_{-}x_{+}x_{c}\left( x_{-}+x_{+}+x_{c}\right) ,
\label{alRelation}
\end{equation}%
\begin{equation}
\lambda =\frac{2}{\left( x_{-}+x_{+}\right) \left( x_{-}+x_{c}\right) \left(
x_{+}+x_{c}\right) },  \label{lxRelation}
\end{equation}%
\begin{equation}
x_{0}=-\left( x_{-}+x_{+}+x_{c}\right) ,
\end{equation}%
where $x_{-}$, $x_{+}$, and $x_{c}$\ are considered as three fundamental
parameters of spacetime. Similar to the Kerr case, there should be a maximum
value for the rotation parameter, say $\alpha _{\max }$, such that we have
black holes in the range $0\leq \alpha ^{2}\leq \alpha _{\max }^{2}$ and a
naked singularity for $\alpha ^{2}>\alpha _{\max }^{2}$. By considering (\ref%
{alRelation})-(\ref{lxRelation}),\ and also, the condition $%
x_{-}<x_{+}<x_{c} $ between roots,\ we find that the maximum $\alpha _{\max
} $ happens whenever $x_{-}\lesssim x_{+}\lesssim x_{c}$, hence all the
horizons are closely spaced for $\alpha _{\max }$. Therefore, we should take
into account the approximation $x_{-}\approx x_{+}\approx x_{c}$\ to obtain $%
\alpha _{\max }$. On the other hand, $0\leq \alpha ^{2}\leq \alpha _{\max
}^{2}$\ and (\ref{alRelation})\ show that the cosmological constant should
obey the following interval as well%
\begin{equation}
0\leq \lambda \leq \lambda _{crit},\quad \lambda _{crit}=\frac{\alpha _{\max
}^{2}}{x_{-}x_{+}x_{c}\left( x_{-}+x_{+}+x_{c}\right) }.
\end{equation}%
\ 

Note that in order to have black holes, there is always a maximum
cosmological constant, say $\lambda _{\max }$, which corresponds to an
arbitrary rotation parameter $\alpha $ in the range $0\leq \alpha \leq
\alpha _{\max }$. The maximum value of $\lambda $,\ in the general case,\ is
corresponding to $\alpha _{\max }$ which is denoted as critical cosmological
constant $\lambda _{crit}$\ in the aforementioned inequality. Therefore,
generally speaking, there is an interval for $\lambda _{\max }$ that depends
on the rotation parameter $\alpha $, as $\lambda _{\max }^{(SdS)}(\alpha
=0)\leq \lambda _{\max }\leq \lambda _{crit}\left( \alpha =\alpha _{\max
}\right) $\ so that its lower bound represents the maximum value of $\lambda 
$\ for the SdS black hole and $\lambda \approx \lambda _{\max }^{(SdS)}$\
characterizes the near-extremal SdS solution. In other words, there is a
maximum value of $\lambda $\ for an arbitrary rotation parameter $\alpha $,
and similarly, there is a maximum value of $\alpha $\ for an arbitrary
cosmological constant $\lambda $ (for example, for the SdS black hole with $%
\alpha =0$, the cosmological constant ranges within $0\leq \lambda \leq
\lambda _{\max }^{(SdS)}$).

Now, in order to obtain $\alpha _{\max }$, and then $\lambda _{\max
}^{(SdS)} $\ and $\lambda _{crit}$, we need to take into account the
condition $x_{-}\lesssim x_{+}\lesssim x_{c}$, as we mentioned above. To do
so, we first equate Eqs. (\ref{DeltaFun})\ and (\ref{DeltaZeros}) while
considering $x_{+}$\ and $x_{c}$\ as two independent variables to obtain the
following relations%
\begin{equation}
\alpha =\sqrt{\frac{4x_{+}x_{c}+\Pi -\Upsilon }{2\left( x_{+}+x_{c}\right) }}%
,  \label{aOrginal}
\end{equation}%
\begin{eqnarray}
x_{-} &=&\frac{1}{2\sqrt{\left( x_{c}+x_{+}-2\right) \left(
x_{c}+x_{+}\right) }}\left\{ x_{c}^{4}+x_{+}^{4}+4\Pi +2\Upsilon \right. 
\notag \\
&&\left. +2x_{+}x_{c}\left[ 2\left( x_{+}+x_{c}+1\right) -x_{c}x_{+}\right]
\right\} ^{\frac{1}{2}}  \notag \\
&&-\frac{1}{2}\left( x_{+}+x_{c}\right) ,  \label{InnerHorOrginal}
\end{eqnarray}%
\begin{equation}
x_{0}=-\left( x_{-}+x_{+}+x_{c}\right) ,
\end{equation}%
\begin{equation}
\lambda =\frac{-\Pi +\Upsilon }{2x_{+}^{2}x_{c}^{2}\left( x_{+}+x_{c}\right) 
},  \label{lOrginal}
\end{equation}%
with $\Pi $\ and $\Upsilon $\ being%
\begin{equation}
\Pi =\sqrt{\Upsilon ^{2}-4x_{+}^{2}x_{c}^{2}\left( x_{+}+x_{c}-2\right)
\left( x_{+}+x_{c}\right) },
\end{equation}%
\begin{equation}
\Upsilon =x_{+}^{3}+x_{+}x_{c}\left( x_{+}+2\right) +x_{c}^{2}\left(
x_{+}+x_{c}\right) .
\end{equation}

Then, we take into account the near-extremal regime $x_{c}\rightarrow x_{+}$
in the above-mentioned formulas, i.e. when the cosmological horizon $x_{c}$
is very close to the black hole event horizon $x_{+}$\ ($x_{c}-x_{+}<<x_{+}$%
). Hence, we obtain the following relations for $\alpha $, $x_{-}$, $x_{0}$,
and $\lambda $ in the nearly extreme regime 
\begin{equation}
\alpha \approx \sqrt{\frac{x_{+}}{2}}\left( 1-2x_{+}+\sqrt{1+8x_{+}}\right)
^{\frac{1}{2}},  \label{aNearEx}
\end{equation}%
\begin{equation}
x_{-}\approx -x_{+}+\sqrt{\frac{x_{+}}{2\left( x_{+}-1\right) }}\left(
1+2x_{+}+\sqrt{1+8x_{+}}\right) ^{\frac{1}{2}},  \label{InnerHorNearEx}
\end{equation}%
\begin{equation}
x_{0}\approx -2x_{+}-x_{-},
\end{equation}%
\begin{equation}
\lambda _{\max }\approx \frac{1+2x_{+}-\sqrt{1+8x_{+}}}{2x_{+}^{3}},
\label{lNearEx}
\end{equation}%
where we replaced $\lambda $\ with $\lambda _{\max }$\ since $x_{c}\approx
x_{+}$. Now, the maximum value of the rotation parameter, $\alpha _{\max }$,
can be obtained by taking the limit $x_{+}\rightarrow x_{-}$\ in the
aforementioned relations. By considering $x_{-}\approx x_{+}$\ in (\ref%
{InnerHorNearEx}), we obtain a maximum value for the event horizon as below%
\begin{equation}
x_{+}=\frac{3+2\sqrt{3}}{4},  \label{MaxEH}
\end{equation}%
that is indeed a coincidental point for all the three horizons ($%
x_{c}\approx x_{+}\approx x_{-}\approx \left( 3+2\sqrt{3}\right) /4$), hence
it maximizes the rotation parameter in Eq. (\ref{aNearEx}). Therefore, by
substituting (\ref{MaxEH})\ in (\ref{aNearEx}), we obtain%
\begin{equation}
\alpha _{\max }=\frac{1}{4}\left( 9+6\sqrt{3}\right) ^{\frac{^{1}}{2}%
}\approx 1.101,  \label{1.101}
\end{equation}%
that is slightly higher than unity for the standard Kerr black hole ($\alpha
_{\max }>\alpha _{\max }^{Kerr}=1$). It is worthwhile to mention that this
value corresponds to the maximum possible value of the cosmological
constant, $\lambda _{crit}\left( \alpha =\alpha _{\max }\right) $, and
therefore, it will be less for lower values of the cosmological constant.
Thus, the rotation parameter ranges%
\begin{equation}
0\leq \alpha \leq \alpha _{\max }
\end{equation}
with $\alpha _{\max }$ given by (\ref{1.101}).

%
%
%
%
%
%%%%%%%%%%%%%%%%%%%%%%%%%%% KdS valid area %%%%%%%%%%%%%%%%%%%%%%%%%%
\begin{figure}[tbp]
\centering
\includegraphics[width=0.45\textwidth]{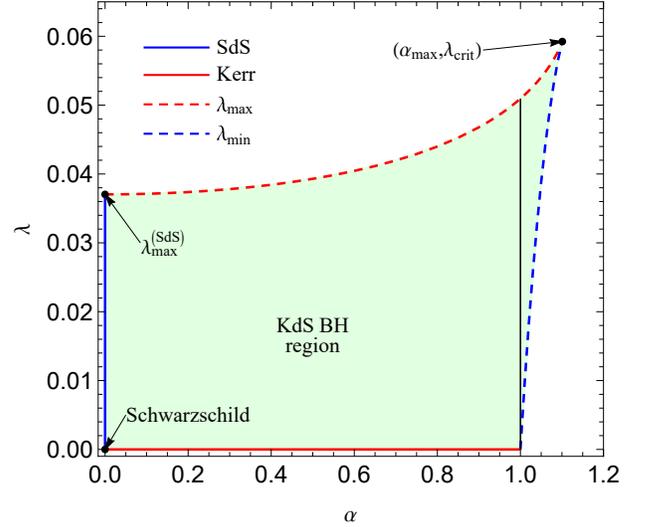}
\caption{The valid area of KdS black holes in $\protect\alpha -\protect%
\lambda $\ parameter space. The shaded green region belongs to KdS black
holes while the marginal points on the left indicate SdS black holes and on
the bottom represent standard Kerr solutions. This figure also shows how a
background cosmological constant extends the parameter space.}
\label{KdSarea}
\end{figure}
%%%%%%%%%%%%%%%%%%%%%%%%%%%%%%%%%%%%%%%%%%%%%%%%%%%%%%%%%%%%%%%
Now, we obtain bounds on maximum values of the cosmological constant $%
\lambda _{\max }$, namely $\lambda _{\max }^{(SdS)}(\alpha =0)$\ and $%
\lambda _{crit}\left( \alpha =\alpha _{\max }\right) $. The maximum value
for the event horizon is given in Eq. (\ref{MaxEH})\ that corresponds to
maximally rotating black holes with $\alpha =\alpha _{\max }$. Therefore, by
substituting Eq. (\ref{MaxEH}) in Eq. (\ref{lNearEx}), we obtain $\lambda
_{crit}\left( \alpha =\alpha _{\max }\right) =16/\left( 3+2\sqrt{3}\right)
^{3}$.

On the other hand, since SdS is static $\alpha =0$, we set $x_{-}=0$\ in (%
\ref{InnerHorNearEx}) and obtain the maximum value of the event horizon as $%
x_{+}=3\equiv x_{+\max }^{(SdS)}$\ for this case (we set $x_{-}=0$ because
SdS black holes have no inner horizon). By replacing this value in Eq. (\ref%
{lNearEx}), one can find the maximum value of the cosmological constant for
the static case as $\lambda _{\max }\left( \alpha =0\right) =1/27\equiv
\lambda _{\max }^{(SdS)}$. We summarized the results of this section on the
bounds of KdS parameters as follows%
\begin{equation}
\left\{ 
\begin{array}{c}
0\leq \alpha \leq \alpha _{\max }, \\ 
\\ 
\lambda _{\max }^{(SdS)}\leq \lambda _{\max }\leq \lambda _{crit}, \\ 
\\ 
x_{+\max }^{(SdS)}\leq x_{+\max }\leq x_{+crit},%
\end{array}%
\right.  \label{restrictions}
\end{equation}%
with%
\begin{equation}
\left\{ 
\begin{array}{c}
\alpha _{\max }=\frac{1}{4}\left( 9+6\sqrt{3}\right) ^{\frac{^{1}}{2}}, \\ 
\\ 
\lambda _{\max }^{(SdS)}=\frac{1}{27},\quad \lambda _{crit}=\frac{16}{\left(
3+2\sqrt{3}\right) ^{3}}, \\ 
\\ 
x_{+\max }^{(SdS)}=3,\quad x_{+crit}=\frac{3+2\sqrt{3}}{4}.%
\end{array}%
\right.
\end{equation}

Therefore, the maximum value of the cosmological constant in the Kerr
geometry must be in the aforementioned interval. For instance, the relation
of $\lambda $\ given in (\ref{lxRelation}) must obey $0\leq \lambda \leq
\lambda _{\max }^{(SdS)}$ for the static case $\alpha =0$,\ while $\lambda
=0 $\ represents the standard Schwarzschild solution and $\lambda \approx
\lambda _{\max }^{(SdS)}$\ denotes nearly extreme SdS black holes. Other
cases far from these two extreme values and within this range are known as
SdS black holes while we have a naked singularity for $\lambda >\lambda
_{\max }^{(SdS)}$.

On the other hand, one may note that for an arbitrary value of the rotation
parameter in the range $1<\bar{\alpha}\leq \alpha _{\max }$, $\lambda $\
acquires a minimum value as well and must obey $\lambda _{\min }\leq \lambda
\leq \lambda _{\max }$ for a given $\bar{\alpha}$. Obtaining these bounds on
the parameters of KdS black holes is important since they will help us to
find valid values of the redshifted photons emitted by massive particles
orbiting a KdS black hole.

Various regions of KdS black holes in the $\alpha -\lambda $\ plane are
illustrated in Fig. \ref{KdSarea}. In this figure, the continuous vertical
blue line $\alpha =0$\ represents SdS black holes, the continuous horizontal
red line $\lambda =0$\ shows the standard Kerr black holes, and there are
standard Schwarzschild solutions where they join $\left\{ \alpha =0,\lambda
=0\right\} $.\ Extreme points $\left\{ \alpha =0,\lambda =\lambda _{\max
}^{(SdS)}\right\} $\ and $\left\{ \alpha =\alpha _{\max },\lambda =\lambda
_{crit}\right\} $ that we have obtained analytically in (\ref{restrictions})
are shown on the top corners. To obtain the $\lambda _{\max }$-dashed line\
on the top of the shaded green area, we used Eqs. (\ref{aNearEx}) and (\ref%
{lNearEx}) while one can employ the relations (\ref{aOrginal})-(\ref%
{lOrginal}) in order to find the $\lambda _{\min }$-dashed line on the right
of the shaded green area assuming $x_{-}=x_{+}$. Thus, we derived all the
marginal points of KdS black holes in the parameter space $\alpha -\lambda $ 
\textit{analytically} that are presented in Eqs. (\ref{aOrginal})-(\ref%
{lOrginal}), (\ref{aNearEx})-(\ref{lNearEx}), and (\ref{restrictions}). Note
that some of these bounds have been found in \cite{KdSECO}\ from a different
approach.

It is worthwhile to mention that we have a naked singularity for $\lambda
>\lambda _{\max }$, whereas for $\alpha >\alpha _{\max }$, the inner and
outer horizons vanish and there is just a cosmological horizon assuming $%
\lambda \neq 0$. One can see that the non-vanishing cosmological constant
introduces two important modifications to the standard Kerr geometry: (i) It
leads to a new horizon, known as the cosmological horizon, and (ii) allows
having higher values for the rotation parameter. As we shall see in Sec. \ref%
{CircularEmitters}, the cosmological constant also modifies the particles'
motion and leads to an upper bound on the radius of stable emitters in
circular motion.

\subsection{Geodesics of timelike and null particles in the Kerr-dS
background}

\label{Geodesics}

The equation of motion of test massless/massive particles in the rotating
spacetimes is described by the geodesic equations. In this regard, the
geodesic equations can be obtained by using the separation of variables of
the Hamilton-Jacobi equation. The Hamilton-Jacobi equation, for a given
background $g_{\mu \nu }\left( x^{\rho }\right) $, leading to the geodesic
equations can be written as \cite{CarterCons}%
\begin{equation}
2\frac{\partial S}{\partial \tau }=-g^{\mu \nu }\frac{\partial S}{\partial
x^{\mu }}\frac{\partial S}{\partial x^{\nu }},  \label{HamiltonJacobi}
\end{equation}%
where $S$ represents Hamilton principal function. In this relation, $\tau $
is the proper time that parametrizes the particle worldline and is related
to the affine parameter $\sigma $ by $\tau =\sigma m$ which $m$\ is the
particle rest mass ($\tau $ represents the affine parameter in the case of
photons). For the KdS spacetime, the Hamilton principal function $S$ can be
separated as%
\begin{equation}
S=\frac{1}{2}\eta \tau -\bar{E}t+\bar{L}\varphi +S_{r}\left( r\right)
+S_{\theta }\left( \theta \right) ,  \label{PrincipalFunction}
\end{equation}%
for both timelike ($\eta =1$) and null ($\eta =0$) particles, and $%
S_{r}\left( r\right) $ is a function of $r$\ while $S_{\theta }\left( \theta
\right) $ is a function of $\theta $ only. The constants of motion $\bar{E}$
and $\bar{L}$, respectively, correspond to conserved energy $E_{0}$ and
angular momentum $L_{0}$ of massive particles obtained through the following
equations%
\begin{equation}
\bar{E}=\frac{E_{0}}{m}=-g_{\mu \nu }\xi ^{\mu }U^{\nu },
\label{EnergyParticle}
\end{equation}%
\begin{equation}
\bar{L}=\frac{L_{0}}{m}=g_{\mu \nu }\psi ^{\mu }U^{\nu },
\label{MomentumParticle}
\end{equation}%
where $\xi ^{\mu }=\delta _{t}^{\mu }$\ is the timelike Killing vector field
and $\psi ^{\mu }=\delta _{\varphi }^{\mu }$\ is the rotational Killing
vector field of the spacetime, and $U^{\mu }$ is the $4$-velocity of
particles which is normalized to unity $U^{\mu }U_{\mu }=-1$.

On the other hand, by substituting the decomposition (\ref{PrincipalFunction}%
) into the Hamilton-Jacobi equation (\ref{HamiltonJacobi}), we get the
following equality%
\begin{eqnarray}
&&\eta a^{2}\cos ^{2}\theta +\Delta _{\theta }\left( \frac{dS_{\theta
}\left( \theta \right) }{d\theta }\right) ^{2}  \notag \\
&&+\frac{\Xi ^{2}}{\Delta _{\theta }\sin ^{2}\theta }\left( a\bar{E}\sin
^{2}\theta -\bar{L}\right) ^{2}  \notag \\
&=&-\eta r^{2}-\Delta _{r}\left( \frac{dS_{r}\left( r\right) }{dr}\right)
^{2}  \notag \\
&&+\frac{\Xi ^{2}}{\Delta _{r}}\left[ \left( r^{2}+a^{2}\right) \bar{E}-a%
\bar{L}\right] ^{2},  \label{equality}
\end{eqnarray}%
where the lhs is a function of $\theta $\ only and the right-hand side (rhs)
just depends on the $r$-coordinate. Therefore, either side is equal to a
constant of motion, known as the Carter constant $\mathcal{C}$\ with the
following form \cite{CarterCons}%
\begin{equation}
\mathcal{C}=\mathcal{K}+\left( \bar{L}-a\bar{E}\right) ^{2}\Xi ^{2},
\label{CarterConstant}
\end{equation}%
which $\mathcal{K}$ is a constant that arises from the contraction of the
Killing tensor field $K_{\mu \nu }$\ of Kerr-dS spacetime with the $4$%
-velocity as $\mathcal{K}=K_{\mu \nu }U^{\mu }U^{\nu }$. Now, by making use
of (\ref{EnergyParticle})--%, (\ref{MomentumParticle}), (\ref{equality}), 
(\ref{CarterConstant}), and the unity condition $U^{\mu }U_{\mu }=-1$, we
can obtain the $4$-velocity components of massive particles ($\eta =1$) in
terms of constants of motion $\bar{E}$, $\bar{L}$, and $\mathcal{K}$ as
follows%
\begin{eqnarray}
U^{t} &=&\frac{\Xi ^{2}}{\Sigma \Delta _{\theta }\Delta _{r}}\left\{ a\left[
\Delta _{r}-\left( a^{2}+r^{2}\right) \Delta _{\theta }\right] \bar{L}%
\right.   \notag \\
&&\left. +\left[ \left( a^{2}+r^{2}\right) ^{2}\Delta _{\theta }-a^{2}\sin
^{2}\theta \Delta _{r}\right] \bar{E}\right\} ,  \label{Ut}
\end{eqnarray}%
\begin{eqnarray}
\Sigma ^{2}\left( U^{r}\right) ^{2} &=&\Xi ^{2}\left[ \left(
a^{2}+r^{2}\right) \bar{E}-a\bar{L}\right] ^{2}  \notag \\
&&-\Delta _{r}\left[ \mathcal{K}+r^{2}+\Xi ^{2}(\bar{L}-a\bar{E})^{2}\right] 
\notag \\
&\equiv &V_{r}\left( r\right) ,  \label{Ur}
\end{eqnarray}%
\begin{eqnarray}
\Sigma ^{2}\left( U^{\theta }\right) ^{2} &=&\Delta _{\theta }\left( 
\mathcal{K}-a^{2}\cos ^{2}\theta \right) -a^{2}\Xi ^{2}\left( \sin
^{2}\theta -\Delta _{\theta }\right) \bar{E}^{2}  \notag \\
&&-\Xi ^{2}\left( \frac{1}{\sin ^{2}\theta }-\Delta _{\theta }\right) \bar{L}%
^{2}+2a\Xi ^{2}\left( 1-\Delta _{\theta }\right) \bar{E}\bar{L}  \notag \\
&\equiv &V_{\theta }\left( \theta \right) ,  \label{Uth}
\end{eqnarray}%
\begin{eqnarray}
U^{\varphi } &=&\frac{\Xi ^{2}}{\Sigma \Delta _{\theta }\Delta _{r}\sin
^{2}\theta }\left\{ \left( \Delta _{r}-a^{2}\Delta _{\theta }\sin ^{2}\theta
\right) \bar{L}\right.   \notag \\
&&\left. +a\sin ^{2}\theta \left[ \left( a^{2}+r^{2}\right) \Delta _{\theta
}-\Delta _{r}\right] \bar{E}\right\} ,  \label{Uph}
\end{eqnarray}%
where the rhs of Eq. (\ref{Ur})\ is a function of $r$\ and the rhs of Eq. (%
\ref{Uth}) is a function of $\theta $ only. Note that Eqs. (\ref%
{EnergyParticle}) and (\ref{MomentumParticle}) have been used to obtain Eqs.
(\ref{Ut}) and (\ref{Uph}), while Eqs. (\ref{equality}), (\ref%
{CarterConstant}), and the unity condition $U^{\mu }U_{\mu }=-1$ have been
employed to get the relations (\ref{Ur}) and (\ref{Uth}).

From Eq. (\ref{Uth}), it is clear that the constant of motion $\mathcal{K}$
[that is related to the Carter constant by Eq. (\ref{CarterConstant})]
represents a measure of how much the geodesic of particles deviates from the
equatorial plane $\theta =\pi /2$, where this constant vanishes. Therefore,
the test particles moving in the equatorial plane have zero $\mathcal{K}$,
whereas it is non-vanishing whenever particles cross the equatorial plane.

The first-order differential equations presented in Eqs. (\ref{Ut})-(\ref%
{Uph})\ show the geodesic equations of massive particles for every direction
in the KdS background in terms of the constants of motion $\bar{E}$, $\bar{L}
$, and $\mathcal{K}$. These equations reduce to the corresponding relations
given in \cite{PRDhn} for Kerr geometry in the limit $\Lambda \rightarrow 0$%
, as it should be. Therefore, the cosmological constant encodes deviations
of KdS black holes from the standard Kerr background.

On the other hand, a similar strategy can be followed to obtain the null
geodesics of photons with $4$-momentum $k^{\mu }$ moving between the event
horizon and cosmological horizon of the KdS spacetime. For massless test
particles, the conserved energy $\bar{E}_{\gamma }$ and angular momentum $%
\bar{L}_{\gamma }$ of particles can be found through the following relations%
\begin{equation}
\bar{E}_{\gamma }=-g_{\mu \nu }\xi ^{\mu }k^{\nu },  \label{EnergyPhoton}
\end{equation}%
\begin{equation}
\bar{L}_{\gamma }=g_{\mu \nu }\psi ^{\mu }k^{\nu },  \label{MomentumPhoton}
\end{equation}%
where the $4$-momentum $k^{\mu }$ of null particles satisfies $k^{\mu
}k_{\mu }=0$. Besides, the equality (\ref{equality}) takes the form%
\begin{eqnarray}
&&\Delta _{\theta }\left( \frac{dS_{\theta }\left( \theta \right) }{d\theta }%
\right) ^{2}+\frac{\Xi ^{2}}{\Delta _{\theta }\sin ^{2}\theta }\left( a\bar{E%
}_{\gamma }\sin {}^{2}\theta -\bar{L}_{\gamma }\right) ^{2} =  \notag \\
&&-\Delta _{r}\left( \frac{dS_{r}\left( r\right) }{dr}\right) ^{2}+\frac{\Xi
^{2}}{\Delta _{r}}\left[ \left( r^{2}+a^{2}\right) \bar{E}_{\gamma }-a\bar{L}%
_{\gamma }\right] ^{2},  \label{equalityPhoton}
\end{eqnarray}%
for the null particles ($\eta =0$). Now, by making use of Eqs. (\ref%
{EnergyPhoton})-(\ref{equalityPhoton})\ and $k^{\mu }k_{\mu }=0$,\ we obtain
the various components of the $4$-momentum in terms of the constants of
motion $\bar{E}_{\gamma }$, $\bar{L}_{\gamma }$, and $\mathcal{K}_{\gamma }$
as follows%
\begin{eqnarray}
k^{t} &=&\frac{\Xi ^{2}}{\Sigma \Delta _{\theta }\Delta _{r}}\left\{ a\left[
\Delta _{r}-\left( a^{2}+r^{2}\right) \Delta _{\theta }\right] \bar{L}%
_{\gamma }\right.  \notag \\
&&\left. +\left[ \left( a^{2}+r^{2}\right) ^{2}\Delta _{\theta }-a^{2}\sin
^{2}\theta \Delta _{r}\right] \bar{E}_{\gamma }\right\} ,  \label{kt}
\end{eqnarray}%
\begin{eqnarray}
\Sigma ^{2}\left( k^{r}\right) ^{2} &=&\Xi ^{2}\left[ \left(
a^{2}+r^{2}\right) \bar{E}_{\gamma }-a\bar{L}_{\gamma }\right] ^{2}  \notag
\\
&&-\Delta _{r}\left[ \mathcal{K}_{\gamma }+\Xi ^{2}(\bar{L}_{\gamma }-a\bar{E%
}_{\gamma })^{2}\right]  \notag \\
&\equiv &\mathcal{V}_{r}\left( r\right) ,  \label{kR}
\end{eqnarray}%
\begin{eqnarray}
\Sigma ^{2}\left( k^{\theta }\right) ^{2} &=&\Delta _{\theta }\mathcal{K}%
_{\gamma }-a^{2}\Xi ^{2}\left( \sin ^{2}\theta -\Delta _{\theta }\right) 
\bar{E}_{\gamma }^{2}  \notag \\
&&-\Xi ^{2}\left( \frac{1}{\sin ^{2}\theta }-\Delta _{\theta }\right) \bar{L}%
_{\gamma }^{2}+2a\Xi ^{2}\left( 1-\Delta _{\theta }\right) \bar{E}_{\gamma }%
\bar{L}_{\gamma }  \notag \\
&\equiv &\mathcal{V}_{\theta }\left( \theta \right) ,  \label{kth}
\end{eqnarray}%
\begin{eqnarray}
k^{\varphi } &=&\frac{\Xi ^{2}}{\Sigma \Delta _{\theta }\Delta _{r}\sin
^{2}\theta }\left\{ \left( \Delta _{r}-a^{2}\Delta _{\theta }\sin ^{2}\theta
\right) \bar{L}_{\gamma }\right.  \notag \\
&&\left. +a\sin ^{2}\theta \left[ \left( a^{2}+r^{2}\right) \Delta _{\theta
}-\Delta _{r}\right] \bar{E}_{\gamma }\right\} ,  \label{kph}
\end{eqnarray}%
where the rhs of Eq. (\ref{kR})\ is a function of $r$, the rhs of Eq. (\ref%
{kth}) is a function of $\theta $, and $\mathcal{C}_{\gamma }=\mathcal{K}%
_{\gamma }+\left( \bar{L}_{\gamma }-a\bar{E}_{\gamma }\right) ^{2}\Xi ^{2}$\
is the corresponding Carter constant for photons.

It is worthwhile to mention that the equations given in (\ref{Ut})-(\ref{Uph}%
) and (\ref{kt})-(\ref{kph}), respectively, fully describe any geodesic
motion of massive and massless particles in the background of KdS black
holes for given sets of constants of motion $\{\bar{E},$ $\bar{L},\mathcal{K}%
\}$\ and $\{\bar{E}_{\gamma },\bar{L}_{\gamma },\mathcal{K}_{\gamma }\}$.
Hence, these relations govern the most general orbits of massive bodies,
namely nonequatorial elliptic trajectories, and one can obtain arbitrary
particular cases, such as nonequatorial circular orbits, elliptic equatorial
paths, elliptic nonequatorial orbits, nonelliptic trajectories, and
equatorial circular orbits by imposing some suitable boundary conditions.

\subsection{Frequency shift}

\label{FrequencyShift}

In this section, we briefly review our previous results on the frequency
shift of photons emitted by massive particles moving in an axially symmetric
spacetime, a construction based on a general relativistic method \cite%
{PRDhn,PRDbhmn}.

This formalism allows one to express the frequency shift of photons in terms
of orbital parameters of radiant massive objects (stars, for instance), and
the free parameters of the spacetime (the set of parameters $\{M,$ $%
a,\Lambda \}$ in our black hole case study). In this scenario, the probe
particles feel the curvature of spacetime produced by the black hole and
encode the properties of spacetime, characterized by black hole parameters,
in the frequency shift of emitted photons. This capability allows us to
estimate the black hole parameters through measuring the shift in the
frequency of photons and solving an inverse problem.

The orbiting massive particles can emit electromagnetic waves towards us
such that the corresponding photons travel along null geodesics from
emission till detection while the information of the geometry is encoded in
their frequency shift. The frequency of this photon at some position $%
x_{p}^{\mu }=\left( x^{t},x^{r},x^{\theta },x^{\varphi }\right) \mid _{p}$\
reads%
\begin{equation}
\omega _{p}=-\left( k_{\mu }U^{\mu }\right) \mid _{p}\,,  \label{freq}
\end{equation}%
where the index $p$ refers to either the point of emission $x_{e}^{\mu }$ or
detection $x_{d}^{\mu }$ of the photon.

One can see that, in contrast to the commonly used radial velocities in
Newtonian gravity which are coordinate-dependent observables, $\omega _{p}$
is a general relativistic invariant quantity that keeps memory of photons
from emission at $x_{e}^{\mu }$ till detection at $x_{d}^{\mu }$. Therefore,
in the transition from Newtonian gravity to general relativity, it is
logical to take advantage of shifts in the frequency (\ref{freq}) rather
than redshift due to changes in speed. This is because, in addition to the
redshift due to speed changes, the frequency shift due to curvature of
spacetime is also encoded in the observable quantity $\omega _{p}$.

The most general expression for shifts in the frequency $\omega _{p}$ in
axially symmetric backgrounds of the KdS form (\ref{metric}) can be written
as \cite{PRDhn} 
\begin{eqnarray}
1 &+&z_{_{KdS}}\!=\frac{\omega _{e}}{\omega _{d}}  \notag \\
&=&\frac{(E_{\gamma }U^{t}-L_{\gamma }U^{\varphi
}-g_{rr}U^{r}k^{r}-g_{\theta \theta }U^{\theta }k^{\theta })\mid _{e}}{%
(E_{\gamma }U^{t}-L_{\gamma }U^{\varphi }-g_{rr}U^{r}k^{r}-g_{\theta \theta
}U^{\theta }k^{\theta })\mid _{d}}\,,  \label{GeneralShift}
\end{eqnarray}%
where the $4$-velocity $U^{\mu }$ (of emitter/detector) and the $4$-momentum 
$k^{\mu }$ (at emitter/detector position) are given in Eqs. (\ref{Ut})-(\ref%
{Uph}) and Eqs. (\ref{kt})-(\ref{kph}), respectively. Hence, $z_{_{KdS}}$\
is the frequency shift that light signals emitted by massive particles
orbiting a KdS black hole experience in their path along null geodesics
towards a detecting observer. Since we have general forms of $U^{\mu }$\ and 
$k^{\mu }$, the KdS shift (\ref{GeneralShift}) includes arbitrary stable
orbits, such as circular, elliptic, irregular, equatorial, non-equatorial,
etc. Therefore, the frequency shifts of these photons, that are directly
measured observational quantities, along with the orbital parameters of the
emitter and the observer can be used to determine the black hole parameters 
\cite{PRDbhmn}. 
%, something that is not possible in the commonly used radial
%velocities which are coordinate dependent observables.

In the rest of the paper, we shall focus on equatorial circular orbits for
emitters (an important situation describing accretion disks orbiting
supermassive black holes at the core of AGNs and circularly orbiting binary
compact stars) and on radial motion of detectors due to the accelerated
expansion of the Universe produced by the cosmological constant.

\section{Frequency shift in terms of black hole parameters}

\label{frequencyVSparameters}

In Sec. \ref{CircularEmitters}, we obtain the $4$-velocity of emitters in
equatorial circular motion in terms of the black hole parameters. Then, in
Sec. \ref{RadialDetectors}, we express the $4$-velocity of the detector in
radial motion with respect to the emitter-black hole system versus the KdS
parameters. We shall use these results in Sec. \ref{FinalShift} to obtain
the redshift of the KdS black holes $z_{_{KdS}}$ in terms of the parameters
of spacetime for this special configuration and extract the Hubble law from
some physically motivated approximations. Finally, we express the
Schwarzschild mass and the Hubble constant in terms of observational
redshift in Sec. \ref{SchwSec}.

\subsection{Emitters in circular and equatorial orbits}

\label{CircularEmitters}

Usually, the accretion disks orbiting black holes can be well described by
the equatorial circular motion of massive test particles around the rotating
black holes and even any tilted disk should be driven to the equatorial
plane of the rotating background \cite{BardeenPetterson}. Hence, in what
follows, we concentrate our attention on the equatorial circular orbits of
emitters characterized by $\theta =\pi /2$ and $U^{r}=0=U^{\theta }$, to
find the relations between KdS black hole parameters $\{M,a,\Lambda \}$\ and
measured redshifts/blueshifts of light signals detected by an observer
located far away from their source. We also assume that the observer detects
the photons in the equatorial plane $\theta =\pi /2$ since accretion disks
can be detected mostly in an edge-on view from Earth \cite{MCPIII,Darling},
and therefore $k^{\theta }=0$\ identically. With this assumption, we place
the detector in the equatorial plane as well.

At this stage, we express the $4$-velocity $U^{\mu }$ for the equatorial
circular orbits in terms of the KdS black hole parameters $\{M,$ $a,\Lambda
\}$\ in order to substitute in the frequency shift relation (\ref%
{GeneralShift}), hence find a connection between observational
redshift/blueshift and KdS parameters. Therefore, by considering Eqs. (\ref%
{Ut})-(\ref{Uph}), the non-vanishing components $U_{e}^{t}$\ and $%
U_{e}^{\varphi }$ of the emitter read%
\begin{equation}
U_{e}^{t}=\frac{a\left( \Delta _{e}-a^{2}-r_{e}^{2}\right) L_{e}+\left[
\left( a^{2}+r_{e}^{2}\right) ^{2}-a^{2}\Delta _{e}\right] E_{e}}{%
r_{e}^{2}\Delta _{e}}\Xi ,  \label{ut}
\end{equation}%
\begin{equation}
U_{e}^{\varphi }=\frac{\left( \Delta _{e}-a^{2}\right) L_{e}+a\left(
a^{2}+r_{e}^{2}-\Delta _{e}\right) E_{e}}{r_{e}^{2}\Delta _{e}}\Xi ,
\label{uph}
\end{equation}%
where $L_{e}=\Xi \bar{L}_{e}$, $E_{e}=\Xi \bar{E}_{e}$, $\Delta _{e}=\Delta
_{r}(r=r_{e})$, and $r_{e}$\ is the radius of the emitter. In this case, the
Carter constant vanishes whereas the constants of motion $E_{e}$\ and $L_{e}$%
\ can be obtained by taking into account the conditions 
\begin{equation}
V_{r}\left( r\right) =0,\quad \frac{dV_{r}\left( r\right) }{dr}=0,
\end{equation}%
simultaneously for having circular orbits while $V_{r}\left( r\right) $ is
given in Eq. (\ref{Ur}). Therefore,\ one can solve these conditions to get 
\cite{KdSECO}%
\begin{equation}
E_{e}=\frac{r_{e}^{\frac{3}{2}}\left[ 1-\frac{\Lambda }{3}\left(
a^{2}+r_{e}^{2}\right) \right] -2Mr_{e}^{\frac{1}{2}}\pm a\left( M-\frac{%
\Lambda r_{e}^{3}}{3}\right) ^{\frac{1}{2}}}{r_{e}^{\frac{3}{4}}\sqrt{r_{e}^{%
\frac{3}{2}}\left( 1-\frac{\Lambda a^{2}}{3}\right) -3Mr_{e}^{\frac{1}{2}%
}\pm 2a\left( M-\frac{\Lambda r_{e}^{3}}{3}\right) ^{\frac{1}{2}}}},
\label{Energy}
\end{equation}%
\begin{equation}
L_{e}=\frac{\left( a^{2}+r_{e}^{2}\right) \left[ \pm \left( M-\frac{\Lambda
r_{e}^{3}}{3}\right) ^{\frac{1}{2}}-\frac{a\Lambda }{3}r_{e}^{\frac{3}{2}}%
\right] -2aMr_{e}^{\frac{1}{2}}}{r_{e}^{\frac{3}{4}}\sqrt{r_{e}^{\frac{3}{2}%
}\left( 1-\frac{\Lambda a^{2}}{3}\right) -3Mr_{e}^{\frac{1}{2}}\pm 2a\left(
M-\frac{\Lambda r_{e}^{3}}{3}\right) ^{\frac{1}{2}}}},  \label{Momentum}
\end{equation}%
in terms of the KdS black hole parameters while the upper sign corresponds
to a co-rotating object and the lower sign refers to a counter-rotating
object with respect to the angular velocity of the black hole, and we shall
use this convention in the upcoming equations.

Now, by substituting relations (\ref{Energy})\ and (\ref{Momentum})\ into
Eqs.\ (\ref{ut})\ and (\ref{uph}), we can obtain rather simple equations as
follows%
\begin{equation}
U_{e}^{t}\left( r_{e},\pi /2\right) =\frac{r_{e}^{\frac{3}{2}}\pm a\left( M-%
\frac{\Lambda r_{e}^{3}}{3}\right) ^{\frac{1}{2}}}{\mathcal{X}_{\pm }}\Xi ,
\label{UtFinal}
\end{equation}%
\begin{equation}
U_{e}^{\varphi }\left( r_{e},\pi /2\right) =\pm \frac{\left( M-\frac{\Lambda
r_{e}^{3}}{3}\right) ^{\frac{1}{2}}}{\mathcal{X}_{\pm }}\Xi ,
\label{UphFinal}
\end{equation}%
with%
\begin{equation}
\mathcal{X}_{\pm }=r_{e}^{\frac{3}{4}}\sqrt{r_{e}^{\frac{3}{2}}\left( 1-%
\frac{\Lambda a^{2}}{3}\right) -3Mr_{e}^{\frac{1}{2}}\pm 2a\left( M-\frac{%
\Lambda r_{e}^{3}}{3}\right) ^{\frac{1}{2}}}.  \label{Xrelation}
\end{equation}

From Eqs. (\ref{UtFinal})-(\ref{Xrelation}), it is obvious that one should
follow the conditions $3M-\Lambda r_{e}^{3}\geq 0$\ and $\mathcal{X}_{\pm
}^{2}>0$\ in order to have the equatorial circular orbits. The former puts
an upper bound on the emitter radius as $r_{e}^{3}\leq 3M/\Lambda$, a
quantity that must be located within the cosmological horizon. We call this
special distance $\bar{r}=\left( 3M/\Lambda \right) ^{1/3}$ as \textit{zero
gravity radius} (ZGR), a radius where the effective gravity vanishes, as we
shall show below. With the quantities given in (\ref{UtFinal})-(\ref%
{Xrelation}) at hand, we can also obtain the angular velocity of an emitter
orbiting around the KdS black hole in a circular and equatorial orbit as
below%
\begin{equation}
\Omega _{\pm }= \frac{U_{e}^{\varphi }}{U_{e}^{t}}=\pm\frac{ \left( M-\frac{%
\Lambda r_{e}^{3}}{3}\right) ^{\frac{1}{2}}}{r_{e}^{\frac{3}{2}}\pm a\left(
M-\frac{\Lambda r_{e}^{3}}{3}\right) ^{\frac{1}{2}}}.
\end{equation}

%%%%%%%%%%%%%%%%%%%%%%%%%%%%% Stable orbits %%%%%%%%%%%%%%%%%%%
\begin{figure*}[tbp]
\centering
\includegraphics[width=0.32\textwidth]{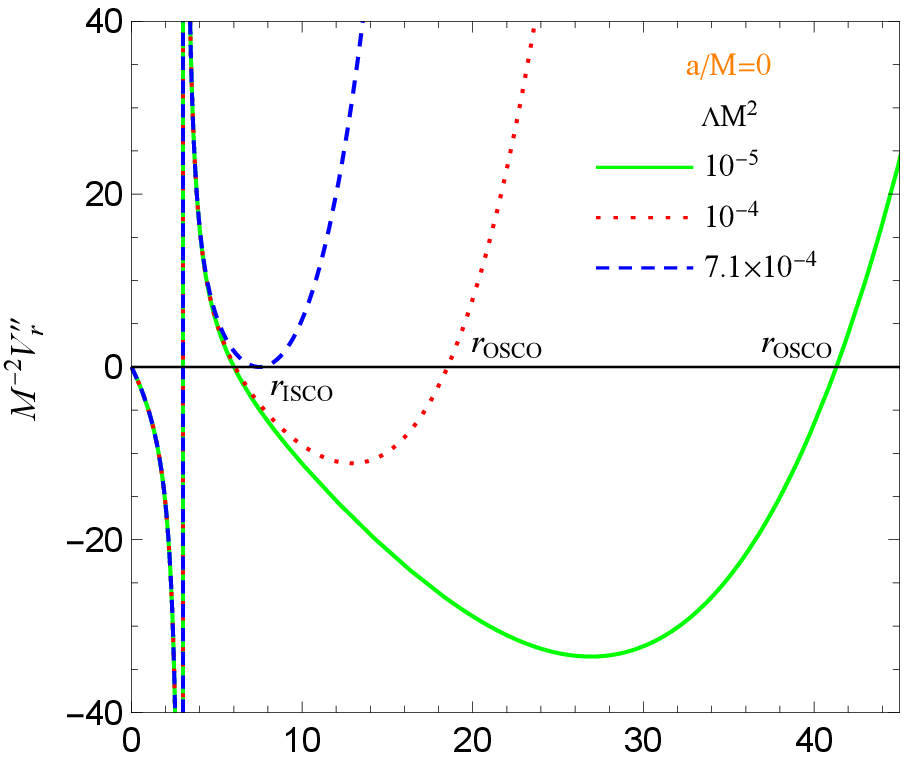} %
\includegraphics[width=0.3\textwidth]{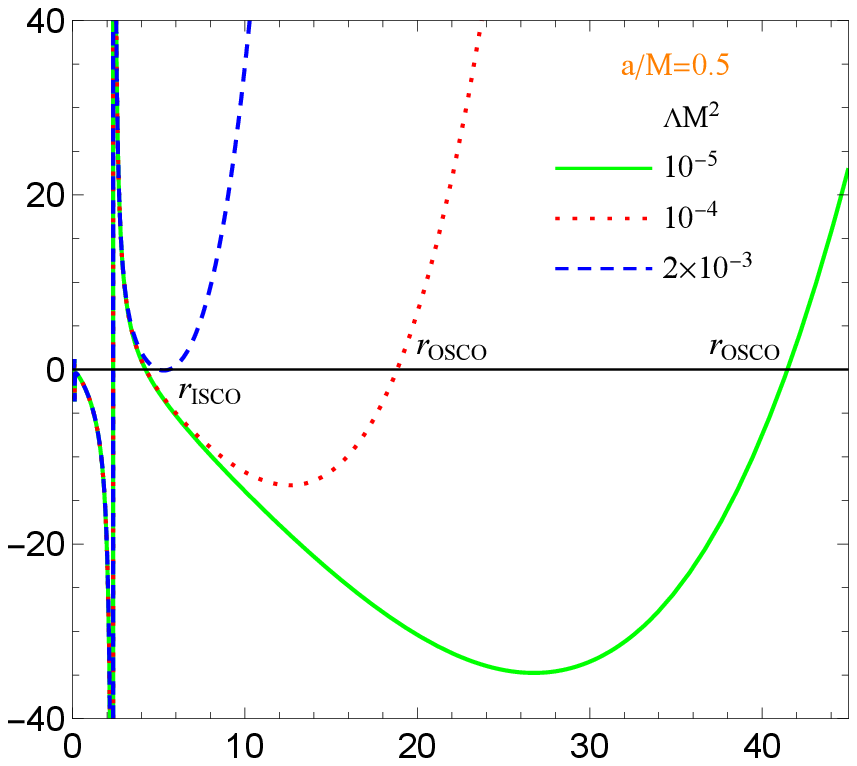} %
\includegraphics[width=0.3\textwidth]{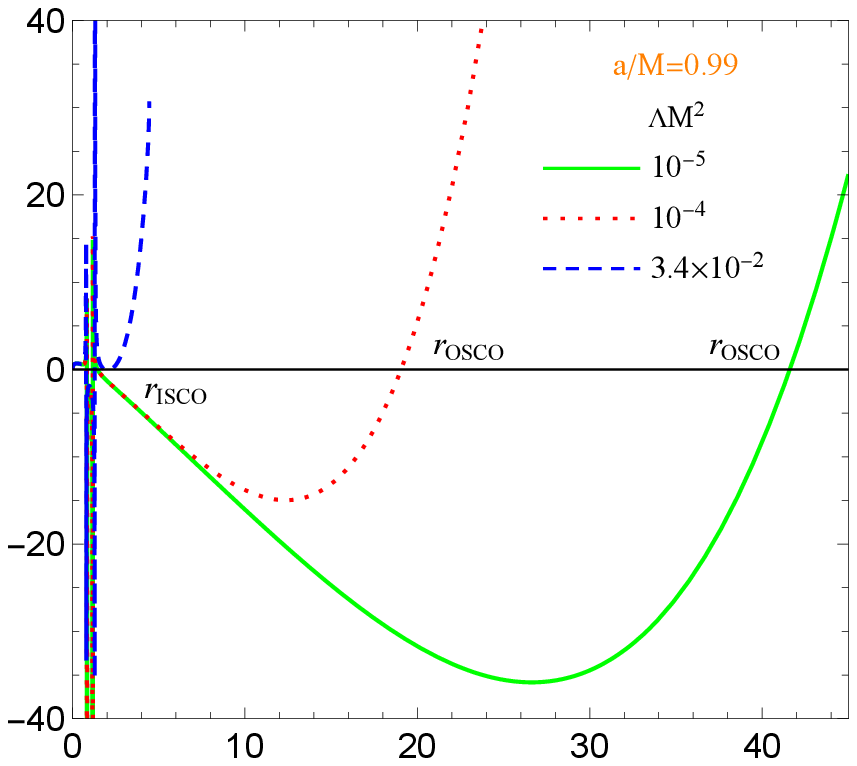} %
\includegraphics[width=0.32\textwidth]{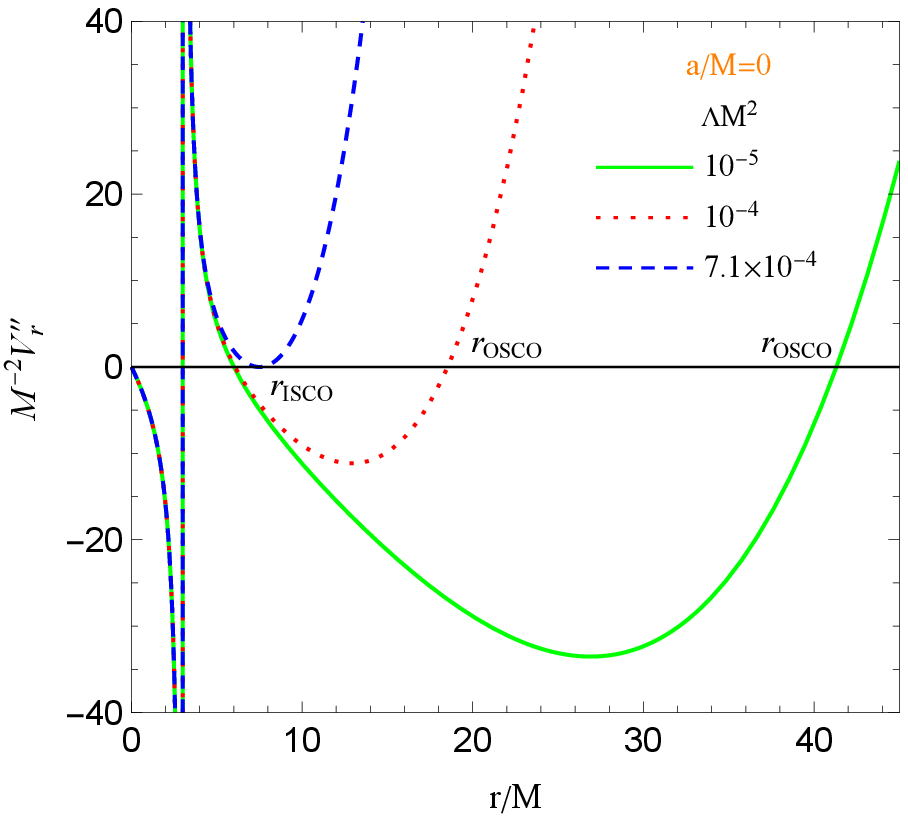} %
\includegraphics[width=0.3\textwidth]{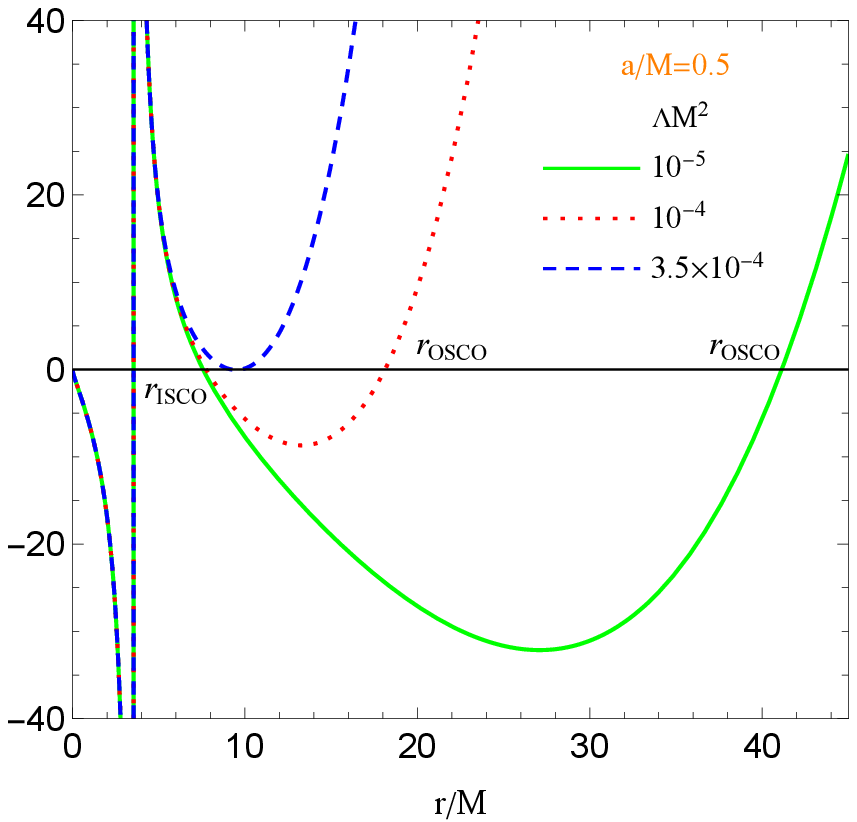} %
\includegraphics[width=0.3\textwidth]{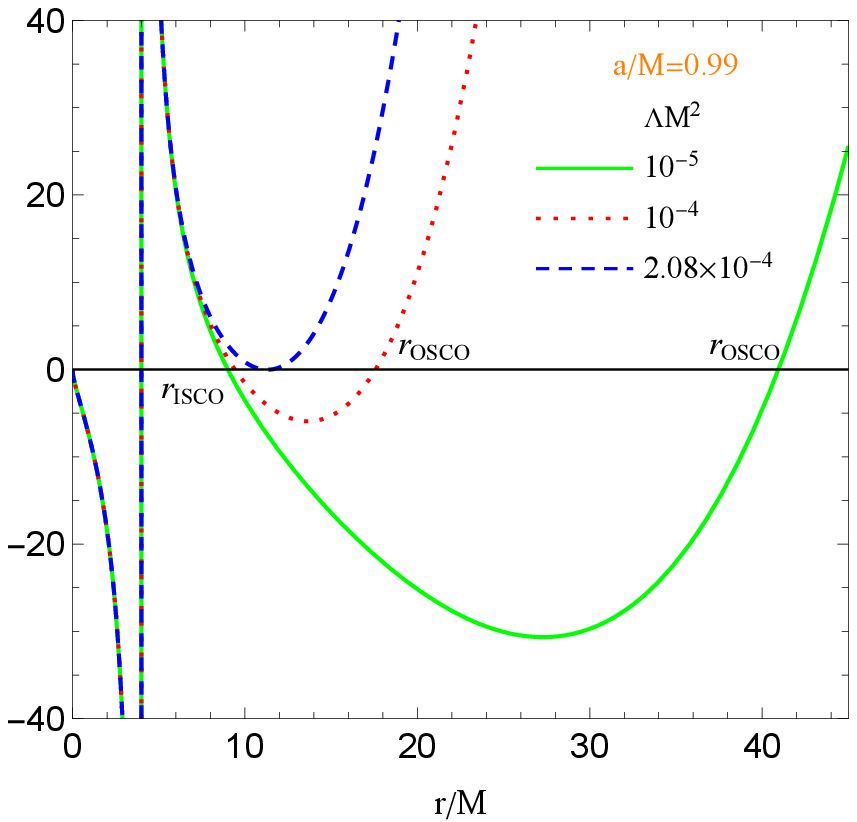}
\caption{The general behavior of $V_{r}^{\prime \prime }$ versus the radial
coordinate for the co-rotating branch (upper panels) and counter-rotating
branch (lower panels). $V_{r}^{\prime \prime }$ is negative between $%
r_{ISCO} $ and $r_{OSCO}$, indicating stable orbits area. $r_{OSCO}$
approaches $r_{ISCO}$ as the cosmological constant increases, and finally,
there will be no stable equatorial circular orbits for sufficiently large $%
\Lambda$.}
\label{sco}
\end{figure*}
%%%%%%%%%%%%%%%%%%%%%%%%%%%%%%%%%%%%%%%%%%%%%%%%%%%%%%%%%%%%%%%

Besides, the non-vanishing components of the $4$-momentum of photons $k^{\mu
}$, given in equations (\ref{kt})-(\ref{kph}), in the equatorial plane
reduce to%
\begin{equation}
k^{t}=\frac{a\left( \Delta _{r}-a^{2}-r^{2}\right) L_{\gamma }+\left[ \left(
a^{2}+r^{2}\right) ^{2}-a^{2}\Delta _{r}\right] E_{\gamma }}{r^{2}\Delta _{r}%
}\Xi ,  \label{ktFinal}
\end{equation}%
\begin{equation}
r^{2}\left( k^{r}\right) ^{2}=\left[ \left( a^{2}+r^{2}\right) E_{\gamma
}-aL_{\gamma }\right] ^{2}-\Delta _{r}(L_{\gamma }-aE_{\gamma })^{2},
\label{krFinal}
\end{equation}%
\begin{equation}
k^{\varphi }=\frac{\left( \Delta _{r}-a^{2}\right) L_{\gamma }+a\left(
a^{2}+r^{2}-\Delta _{r}\right) E_{\gamma }}{r^{2}\Delta _{r}}\Xi ,
\label{kphFinal}
\end{equation}%
with $L_{\gamma }=\Xi \bar{L}_{\gamma }$ and $E_{\gamma }=\Xi \bar{E}%
_{\gamma }$.

On the other hand,\ the condition for having stable orbits in KdS geometry
is given by 
\begin{eqnarray}
V_{r}^{\prime \prime } &\equiv &\frac{d^{2}V_{r}\left( r\right) }{dr^{2}}=-%
\left[ r^{2}+(L-aE)^{2}\right] \Delta _{r}^{\prime \prime }-4r\Delta
_{r}^{\prime }  \notag \\
&&-2\Delta _{r}+4\left( 3r^{2}+a^{2}\right) E^{2}-4aLE\leq 0,  \label{rsco}
\end{eqnarray}%
where prime denotes $\partial _{r}$ and one can use either upper/lower sign
of Eqs. (\ref{Energy})\ and (\ref{Momentum}) to obtain radii of stable
circular orbits of co/counter -rotating stars. However, the polynomial
expression (\ref{rsco})\ is of $10$th-order in $r$\ and could not be solved
analytically, unlike the standard Kerr case.

The general behavior of $V_{r}^{\prime \prime }$\ is illustrated through
Fig. \ref{sco}\ for various values of $\Lambda $\ and $a$ as well as
co/counter -rotating classes. As one can see from this figure, the roots of
relation (\ref{rsco}) characterize the innermost stable circular orbit
(ISCO) $r_{ISCO}$\ and outermost stable circular orbit (OSCO) $r_{OSCO}$\
describing, respectively, the inner edge of orbiting accretion disk and its
outer edge. Therefore, we expect that the lower constraint on the emitter
radius as $r_{e}\geq r_{ISCO}$\ leads to an upper bound on the
redshift/blueshift of orbiting objects, whereas the upper constraint $%
r_{e}\leq r_{OSCO}$\ puts a lower bound on the frequency shift.

Now, it is worthwhile to summarize the most important radii in the KdS
geometry as follows%
\begin{equation}
r_{0}<0<r_{-}<r_{+}<r_{ISCO}<r_{OSCO}<\bar{r}<r_{c},  \label{ImportantRadii}
\end{equation}%
where (i) the inner horizon $r_{-}$, the event horizon $r_{+}$, and the
cosmological horizon $r_{c}$ are the solutions of $\Delta _{r}$\ in Eq. (\ref%
{DeltaR}), (ii) $r_{ISCO}$\ and $r_{OSCO}$ are the solutions of the
stability condition in Eq. (\ref{rsco}), and (iii) the ZGR $\bar{r}=\left(
3M/\Lambda \right) ^{1/3}$ is a maximum radius for having the equatorial
circular orbits obtained through\ (\ref{UtFinal})-(\ref{Xrelation}).

In this study, we are interested in stable orbits satisfying $r_{ISCO}\leq
r_{e}\leq r_{OSCO}$\ for the emitter and far away detectors with the
condition 
\begin{equation}
\bar{r}<r_{d}<r_{c},
\end{equation}%
describing black hole systems in the Hubble flow. However, note that some of
the important radii (\ref{ImportantRadii}) may change/vanish under certain
circumstances, as we discussed in Sec. \ref{KerrDS} (for $r_{-}$, $r_{+}$,
and $r_{c}$) and in Fig. \ref{sco}\ (for $r_{ISCO}$ and $r_{OSCO}$). In
addition, we restrict our calculations to $\Lambda \leq 10^{-4}$\ in order
to have stable orbits, in consistency with Fig. \ref{sco}.

\subsection{Detectors in radial motion}

\label{RadialDetectors}

Here, we should note that the situation for the KdS geometry differs from
the previous cases studied before (see \cite{PRDhn,PRDbhmn,RegularBH,MOG}\
and references therein) and we cannot consider circular orbiting or static
detectors since the behavior of $z_{_{KdS}}$\ in Eq.\ (\ref{GeneralShift})
versus $\Lambda $\ turns out to be unphysical once we take into account the
circular orbits beyond $\bar{r}$. Indeed, because of the accelerated
expansion of the Universe due to the positive cosmological constant at large
scales, the detector should move away from the black hole in the case of
far-away detectors that we are interested in. Therefore, in this case, we
consider a detector that radially moves away from the KdS black hole instead
of usual circularly orbiting or static detectors. This implies that $%
U_{d}^{\varphi }=0=U_{d}^{\theta }$, hence the non-vanishing components of
the $4$-velocity of the detector read [see Eqs. (\ref{Ut})-(\ref{Uph})]%
\begin{equation}
U_{d}^{t}=\frac{\left( a^{2}+r_{d}^{2}\right) ^{2}-a^{2}\Delta _{d}}{%
r_{d}^{2}\Delta _{d}}E_{d}\Xi ,  \label{UtD}
\end{equation}%
\begin{equation}
\left( U_{d}^{r}\right) ^{2}=\frac{\left( a^{2}+r_{d}^{2}\right)
^{2}E_{d}^{2}-\Delta _{d}\left( r_{d}^{2}+a^{2}E_{d}^{2}\right) }{r_{d}^{4}},
\label{UrD}
\end{equation}%
where we have set $L_{d}=0$\ due to the radial motion of the detector, $%
E_{d}=\Xi \bar{E}_{d}$, $\Delta _{d}=\Delta _{r}(r=r_{d})$, and $r_{d}$\ is
the distance between the black hole and the detector. Note that $%
U_{d}^{\varphi }=0$ is just valid for far enough detectors, otherwise the
rotation nature of the spacetime drags the detector, as it can be seen from
Eq. (\ref{Uph}).

As the next step, we need to obtain $E_{d}$\ in terms of the parameters of
the spacetime $\{M,a,\Lambda \}$. One may note that at some radius $r_{d}=R$%
, where the gravitational attraction generated by the black hole mass is
completely balanced by the expansion of the Universe produced by the
cosmological constant, such that $M=M_{\Lambda }$, with $M_{\Lambda }$ being
an effective mass related to the cosmological constant. Thus, the radial
velocity $U_{d}^{r}$ (\ref{UrD}) vanishes at $r_{d}=R$ because the repulsive
nature of the cosmological constant is exactly cancelled by the
gravitational attraction. We obtain the effective mass $M_{\Lambda }$
through the following integral%
\begin{equation}
M_{\Lambda }=\int_{0}^{R}4\pi \rho _{\Lambda }r^{2}dr,
\end{equation}%
where the density of cosmological constant $\rho _{\Lambda }$ is related to
the cosmological constant via $\rho _{\Lambda }=\Lambda /\left( 8\pi
G\right) $ \cite{Carroll}. By performing this integral and equating $%
M=M_{\Lambda }$, we find the vanishing velocity radius as $R=\left(
3M/\Lambda \right) ^{1/3}$ that is exactly equal to the ZGR $\bar{r}$.
Therefore, this is the radius at which the cosmological constant compensates
the gravitational attraction of the black hole, and hence the effective
gravity vanishes as we discussed after Eq. (\ref{Xrelation}). Note that the
angular velocity of the emitter (\ref{UphFinal}) also vanishes for $%
r_{e}=\left( 3M/\Lambda \right) ^{1/3}$ which means the emitter is static at
this point as well.

Now, by replacing $r_{d}=R=\left( 3M/\Lambda \right) ^{1/3}$\ in Eq. (\ref%
{UrD}) and solving $U_{d}^{r}\left( r_{d}=R\right) =0$, one can find the
energy of the detector as below%
\begin{equation}
E_{d}=\left( \frac{3a^{2}\left[ \left( \frac{3M}{\Lambda }\right) ^{1/3}-M%
\right] -9M\left[ \left( \frac{3M}{\Lambda }\right) ^{2/3}-\frac{1}{\Lambda }%
\right] }{a^{2}\left( \frac{3M}{\Lambda }\right) ^{1/3}\left( 3+\Lambda
a^{2}\right) +9M\left( a^{2}+\frac{1}{\Lambda }\right) }\right) ^{\frac{1}{2}%
}.
\end{equation}

In this way, the $4$-velocity components (\ref{UtD})-(\ref{UrD}) can be
written as%
\begin{eqnarray}
U_{d}^{t} &=&\left( \frac{3a^{2}\left[ \left( \frac{3M}{\Lambda }\right)
^{1/3}-M\right] -9M\left[ \left( \frac{3M}{\Lambda }\right) ^{2/3}-\frac{1}{%
\Lambda }\right] }{a^{2}\left( \frac{3M}{\Lambda }\right) ^{1/3}\left(
3+\Lambda a^{2}\right) +9M\left( a^{2}+\frac{1}{\Lambda }\right) }\right) ^{%
\frac{1}{2}}\times  \notag \\
&&\frac{\left( a^{2}+r_{d}^{2}\right) ^{2}-a^{2}\Delta _{d}}{r_{d}^{2}\Delta
_{d}}\Xi ,  \label{utd}
\end{eqnarray}%
\begin{eqnarray}
\left( U_{d}^{r}\right) ^{2} &=&\left( \frac{3a^{2}\left[ \left( \frac{3M}{%
\Lambda }\right) ^{1/3}-M\right] -9M\left[ \left( \frac{3M}{\Lambda }\right)
^{2/3}-\frac{1}{\Lambda }\right] }{a^{2}\left( \frac{3M}{\Lambda }\right)
^{1/3}\left( 3+\Lambda a^{2}\right) +9M\left( a^{2}+\frac{1}{\Lambda }%
\right) }\right) \times  \notag \\
&&\frac{\left( a^{2}+r_{d}^{2}\right) ^{2}-\Delta _{d}a^{2}}{r_{d}^{4}}-%
\frac{\Delta _{d}}{r_{d}^{2}},  \label{urd}
\end{eqnarray}%
in terms of the black hole parameters.

\subsection{Frequency shift versus parameters of spacetime and the Hubble law%
}

\label{FinalShift}

For this configuration, i.e. circularly orbiting emitters and radial motion
of detectors, the general expression for the frequency shift of photons (\ref%
{GeneralShift}) reduces to 
\begin{equation}
1+z_{_{KdS_{1,2}}}\!=\frac{\left. \left( E_{\gamma }U^{t}-L_{\gamma
}U^{\varphi }\right) \right\vert _{e}}{\left. \left( E_{\gamma
}U^{t}-g_{rr}U^{r}k^{r}\right) \right\vert _{d}}=\frac{U_{e}^{t}-b_{e_{(\mp
)}}\,U_{e}^{\varphi }}{U_{d}^{t}-g_{d}U_{d}^{r}\left( \frac{k_{d}^{r}}{%
E_{\gamma }}\right) }\,,  \label{zcircorbits}
\end{equation}%
where we defined the light bending parameter $b$ as $b\equiv L_{\gamma
}/E_{\gamma }$ that represents the deflection of light due to gravitational
field in the vicinity of the KdS black hole. Besides, $g_{d}=g_{rr}\left(
r=r_{d}\right) $ is given in (\ref{grr}) and the ratio $k_{d}^{r}/E_{\gamma
} $\ can be written as (from Eq. (\ref{krFinal}))%
\begin{equation}
\left( \frac{k_{d}^{r}}{E_{\gamma }}\right) ^{2}=\frac{\left[ \left(
a^{2}+r_{d}^{2}\right) -ab_{d_{(\mp )}}\right] ^{2}-\Delta _{d}(b_{d_{(\mp
)}}-a)^{2}}{r_{d}^{2}}.  \label{krOe}
\end{equation}

Note that $b$, presented in Eqs. (\ref{zcircorbits}) and (\ref{krOe}), is
preserved along the whole light path followed by photons from their emission
till their detection due to the fact that $E_{\gamma }$ and $L_{\gamma }$
are constants of motion. Therefore, one can set $b_{e}=b_{d}$ without loss
of generality. Moreover, the subscript $_{(\mp )}$ signs refer to the
deflection of light $b$ at either side of the line of sight, whereas the
subindices $_{_{1}}$ and $_{_{2}}$ in Eq. (\ref{zcircorbits}) correspond to
the $_{_{(-)}}$ and $_{_{(+)}}$ signs, respectively.

On the other hand, the maximum value of the light bending parameter is given
by the condition $k^{r}=0$, where the position vector of orbiting stars with
respect to the black hole center is approximately orthogonal to the line of
sight. Thus, we substitute $k^{t}$\ and $k^{\varphi }$ from Eqs. (\ref%
{EnergyPhoton})-(\ref{MomentumPhoton})\ as well as the condition $k^{r}=0$\
in the photons' equation of motion 
\begin{equation}
k^{\mu }k_{\mu }=0=g_{tt}k^{t}k^{t}+g_{rr}k^{r}k^{r}+g_{\varphi \varphi
}k^{\varphi }k^{\varphi }+2g_{t\varphi }k^{t}k^{\varphi },
\end{equation}%
to find the maximum value of the light bending parameter for the rotating
metric (\ref{metric}) as follows 
\begin{equation}
b_{(\pm )}=-\frac{g_{t\varphi }(\pm )\sqrt{g_{t\varphi
}^{2}-g_{tt}g_{\varphi \varphi }}}{g_{tt}},
\end{equation}%
where in terms of the KdS black hole parameters, we have%
\begin{eqnarray}
b_{(\pm )} &=&\frac{1}{r\left[ 1-\frac{\Lambda }{3}\left( r^{2}+a^{2}\right) %
\right] -2M}\times   \notag \\
&&\left[ -2Ma-\frac{\Lambda }{3}ar\left( r^{2}+a^{2}\right) \right.   \notag
\\
&&\left. \left( \pm \right) r\sqrt{r^{2}+a^{2}-2Mr-\frac{\Lambda r^{2}}{3}%
\left( r^{2}+a^{2}\right) }\right] ,  \label{lbp}
\end{eqnarray}%
for equatorial circular orbits. In this formula, the sign of $b$ denotes the
redshifted and blueshifted photons when their source is co-rotating with
respect to the black hole angular momentum, and vice versa if it is
counter-rotating. In other words, in the frequency shift formulas [like Eq. (%
\ref{zcircorbits})], the minus sign enclosed in parentheses corresponds to
the redshifted photons, whereas the plus sign indicates blueshifted ones.

%%%%%%%%%%%%%%%%%%%%%%%%%%%%% KdS Shift-Lambda %%%%%%%%%%%%%%%%%%%%%%%%%%%%%%%%%%
\begin{figure*}[tbp]
\centering
\includegraphics[width=0.42\textwidth]{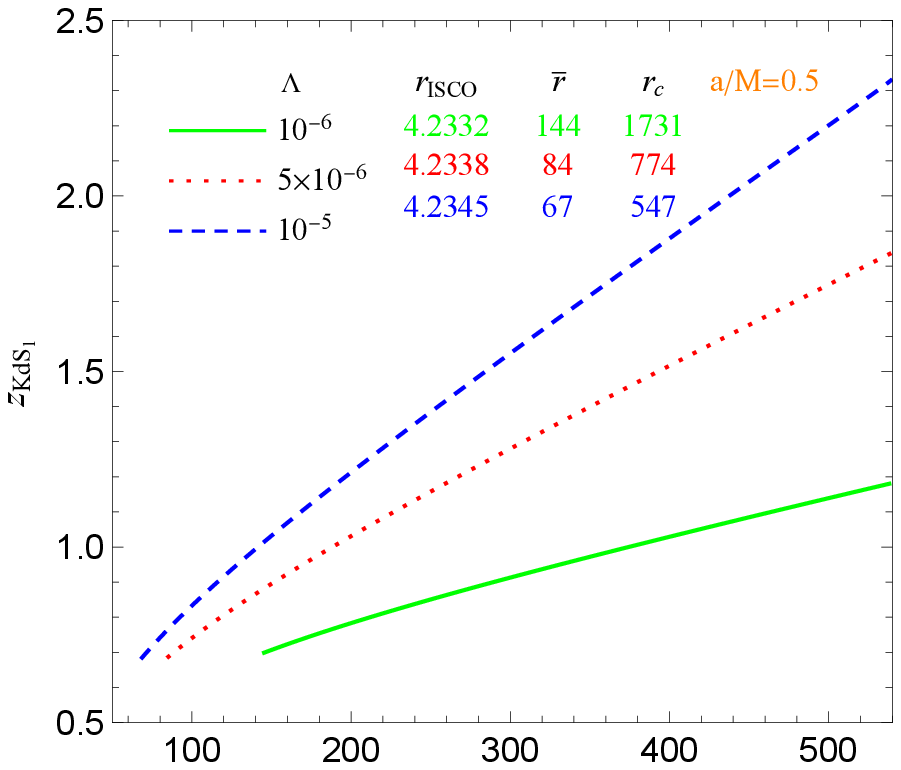} %
\includegraphics[width=0.4\textwidth]{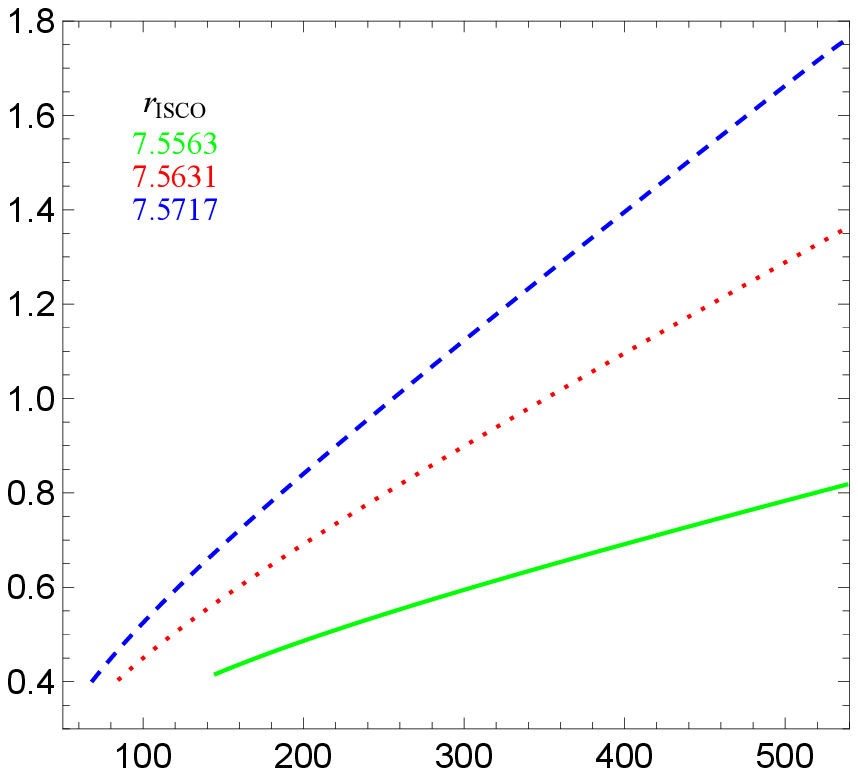} %
\includegraphics[width=0.42\textwidth]{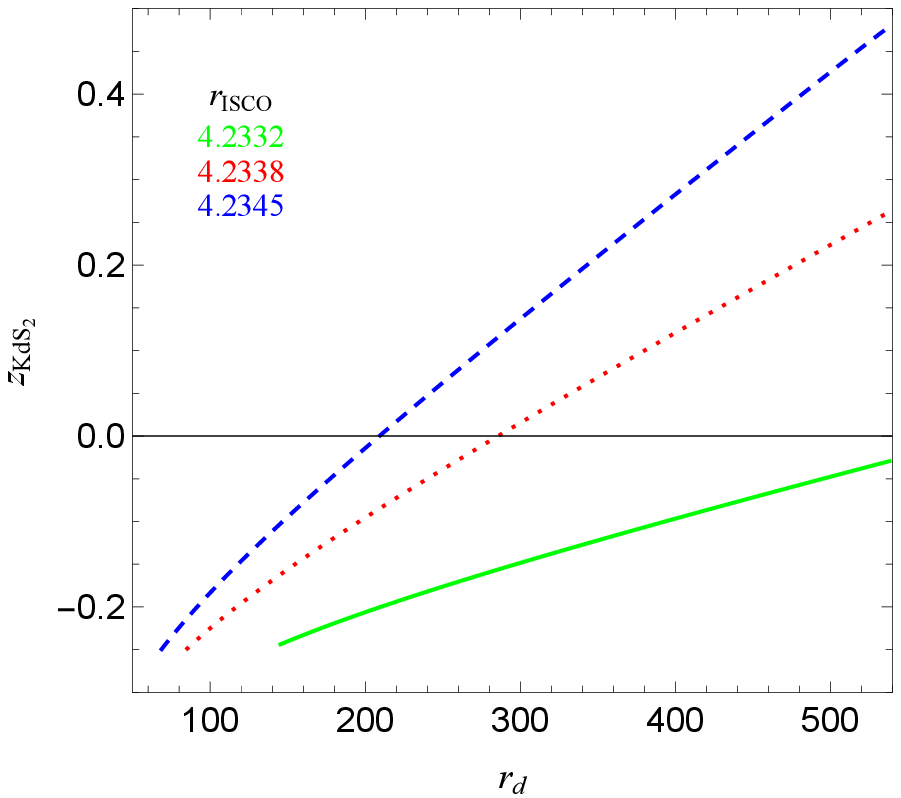} %
\includegraphics[width=0.4\textwidth]{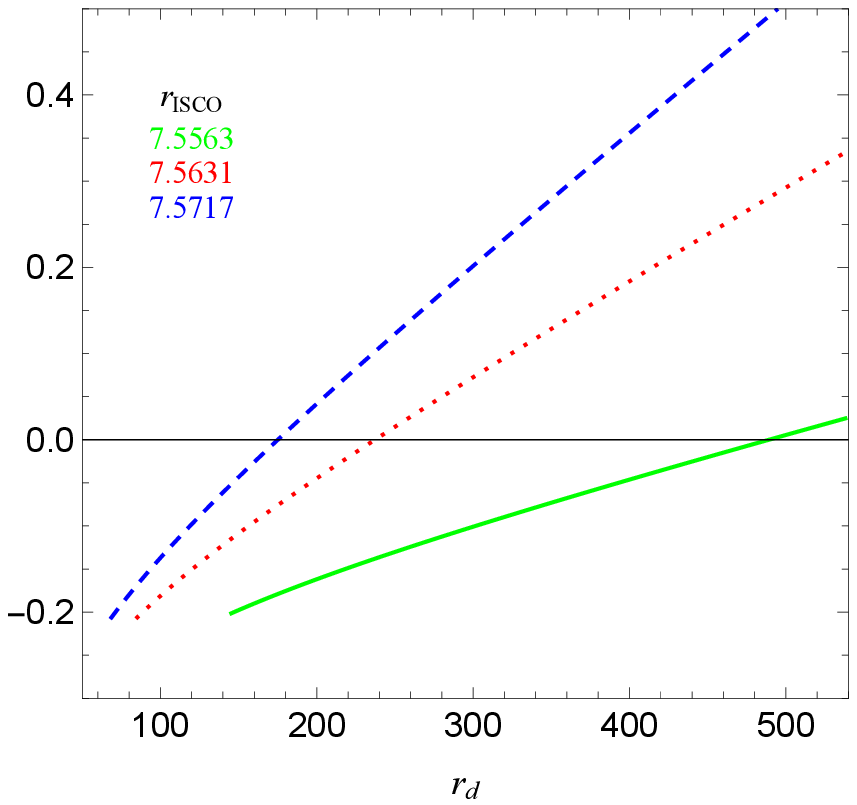}
\caption{The redshift $z_{_{KdS_{1}}}$ and blueshift $z_{_{KdS_{2}}}$ versus
the detector distance $r_{d}$ for the co-rotating branch (left panels) and
counter-rotating branch (right panels). The emitter is orbiting circularly
at the radius $r_{e}=2r_{ISCO}$ for either curve, and we set $M=1$. The
curves start at ZGR $r_{d}=\bar{r}=\left(3M/\Lambda \right) ^{1/3}$ on the
left and terminate before the cosmological horizon $r_{d}<r_{c}$ on the
right.}
\label{KdSLambda}
\end{figure*}
%%%%%%%%%%%%%%%%%%%%%%%%%%%%%%%%%%%%%%%%%%%%%%%%%%%%%%%%%%%%%%%

Now, we can substitute $g_{d}$, $U_{e}^{t}$, $U_{e}^{\varphi }$, $U_{d}^{t}$%
, $U_{d}^{r}$, $k_{d}^{r}/E_{\gamma }$, and $b_{e_{(\mp )}}$, respectivly,
from Eqs. (\ref{grr}), (\ref{UtFinal}), (\ref{UphFinal}), (\ref{utd}), (\ref%
{urd}), (\ref{krOe}), and (\ref{lbp}) into (\ref{zcircorbits}) to find the
final expressions for the redshift $z_{_{KdS_{1}}}$\ and blueshift $%
z_{_{KdS_{2}}}$ of the KdS spacetime as follows 
\begin{equation}
1+z_{_{KdS_{1}}}\!=\frac{r_{e}^{\frac{3}{2}}\pm a\left( M-\frac{\Lambda
r_{e}^{3}}{3}\right) ^{\frac{1}{2}}\pm b_{e_{-}}\,\left( M-\frac{\Lambda
r_{e}^{3}}{3}\right) ^{\frac{1}{2}}}{\Gamma \mathcal{X}_{\pm }}\,,
\label{GeneralRed}
\end{equation}%
\begin{equation}
1+z_{_{KdS_{2}}}\!=1+z_{_{KdS_{1}}}\left\{ b_{e_{-}}\rightarrow
b_{e_{+}}\right\} ,  \label{GeneralBlue}
\end{equation}%
with%
\begin{eqnarray}
\Gamma &=&\frac{1}{r_{d}^{2}\Delta _{d}\Xi }\left\{ \left[ \left(
a^{2}+r_{d}^{2}\right) ^{2}-a^{2}\Delta _{d}\right] E_{d}\Xi \right.  \notag
\\
&&-\sqrt{\left( a^{2}+r_{d}^{2}\right) ^{2}E_{d}^{2}-\Delta _{d}\left(
r_{d}^{2}+a^{2}E_{d}^{2}\right) }\times  \notag \\
&&\left. \sqrt{\left[ \left( a^{2}+r_{d}^{2}\right) -ab_{e_{-}}\right]
^{2}-\Delta _{d}(b_{e_{-}}-a)^{2}}\right\} ,
\end{eqnarray}%
where $b_{e_{-}}=\left. b_{\left( -\right) }\right\vert _{r=r_{e}}$, $%
b_{e_{+}}=\left. b_{\left( +\right) }\right\vert _{r=r_{e}}$, and the upper
(lower) sign refers to a co- (counter-) rotating emitter. Note that the
explicit expressions of $z_{_{KdS_{1,2}}}$ in terms of the black hole
parameters have cumbersome forms, but are not hard to be found since it is a
matter of substituting $\mathcal{X}_{\pm }$, $\Delta _{d}$, $\Xi $, $%
b_{e_{-}}$, and $E_{d}$ in the above-mentioned equations.

The general behavior of\ $z_{_{KdS_{1,2}}}$ versus the detector radius $%
r_{d} $\ for different values of the cosmological constant is illustrated in
Fig. \ref{KdSLambda}. Interestingly, one can see that the shift in frequency
of photons increases as the cosmological constant $\Lambda $ increases. On
the other hand, the farther the detector is from the source, the higher
shift in frequency it observes. This is because the amount of dark energy
between the emitter and detector increases by increasing the distance
leading to larger changes in the frequency. This is what we expected from
the repulsive nature of the cosmological constant, and in this study, the
change in the frequency shift due to this effect is quantified through Eqs. (%
\ref{GeneralRed})-(\ref{GeneralBlue}). This fact leads one to extract the
Hubble law in its original form from $z_{_{KdS_{1,2}}}$\ by taking into
account some physically motivated approximations as we shall show below.

Before deriving the Hubble law, one may note that the frequency shift of KdS
black holes (\ref{zcircorbits}) contains two components including the
gravitational redshift $z_{g}$ as well as kinematic redshifts and blueshifts 
$z_{_{kin_{\pm }}}$\ so that (see \cite{PRDbhmn} for more details)%
\begin{equation}
z_{_{KdS_{1,2}}}=z_{g}+z_{_{kin_{\pm }}}.  \label{ZgZk}
\end{equation}%
\ The gravitational redshift $z_{g}$ of circular motion in the equatorial
plane around the KdS background reads%
\begin{equation}
z_{g}=\frac{U_{e}^{t}-b_{c}\,U_{e}^{\varphi }}{U_{d}^{t}-g_{d}U_{d}^{r}%
\left( \frac{k_{d}^{r}}{E_{\gamma }}\right) },  \label{Zg}
\end{equation}%
where $b_{c}$ is the light bending parameter of a light ray emitted radially
at the central point (on the line of sight). The central light bending $b_{c}
$ is a non-vanishing quantity in rotating spacetimes due to the dragging
effect and can be obtained by considering the condition $k^{\varphi }$\ as
follows%
\begin{equation}
b_{c}=-\frac{g_{t\varphi }}{g_{tt}}=-\frac{6M+\Lambda r_{e}\,\left(
r_{e}^{2}+a^{2}\right) }{3r_{e}-6M-\Lambda r_{e}\,\left(
r_{e}^{2}+a^{2}\right) }a,  \label{bc}
\end{equation}%
where the second equality refers to the equatorial circular orbits around
the KdS black hole. Having the gravitational redshift $z_{g}$ (\ref{Zg})\
and the central light bending parameter $b_{c}$ (\ref{bc})\ at hand, we are
able to quantify all three frequency shift components, namely the KdS shift $%
z_{_{KdS_{1,2}}}$, the gravitational redshift $z_{g}$, and the kinematic
redshift $z_{_{kin_{\pm }}}$\ by considering the relation (\ref{ZgZk}) (note
that $b_{d_{(\mp )}}$ in Eq. (\ref{krOe}) should also be modified to $b_{c}$%
). The importance of these relations becomes more clear when one is going to
quantify the contribution of the gravitational redshift in the vicinity of a
compact object. However, in what follows, we concentrate our attention on
the KdS frequency shift $z_{_{KdS_{1,2}}}$ which is an observable quantity.

It is worth mentioning that for real astrophysical systems consisting of
supermassive black holes orbited by an accretion disk containing photon
sources in the form of water vapor clouds (the so-called megamasers),
usually, the emitter radius is at sub-parsec scale ($r_{e}<1pc$) while the
detector radius is at tens of mega-parsec scale to be within the Hubble flow
($r_{d}>30Mpc$). On the other hand, the mass and angular momentum of the
black holes are of the event horizon radius order $M,a\sim r_{+}$, and the
cosmological constant is of the order $\Lambda \sim 10^{-52}m^{-2}$ \cite%
{Carroll}. Therefore, for a configuration including a supermassive black
hole of the order of $10^{6}$ solar masses in the Hubble flow, we have $%
r_{+}\sim 10^{10}m$, $r_{e}<10^{6}r_{+}$, $r_{d}>10^{13}r_{+}$, and $\Lambda
\sim 10^{-32}r_{+}^{-2}$\ which leads to the following facts 
\begin{eqnarray}
\Lambda a^{2} &\sim &10^{-32};\ \Lambda r_{e}^{2}<10^{-20};\ \frac{M}{r_{d}}%
<10^{-13}, \\
\frac{M}{r_{e}} &>&10^{-6};\ \Lambda r_{d}^{2}>10^{-6}.
\end{eqnarray}

Hence, we can ignore the negligible terms $\left\{ \Lambda a^{2},\Lambda
r_{e}^{2},M/r_{d}\right\} $ and keep dominant terms $\left\{ M/r_{e},\Lambda
r_{d}^{2}\right\} $ for tracking the general relativistic effects. As the
next stage, we expand Eqs. (\ref{GeneralRed})-(\ref{GeneralBlue}) for $%
\left\{ \Lambda r_{e}^{2}\rightarrow 0,\Lambda a^{2}\rightarrow
0,M/r_{d}\rightarrow 0,\Lambda r_{d}^{2}\rightarrow 0\right\} $ and keep the
first dominant term in $\Lambda r_{d}^{2}$, to get 
\begin{equation}
1+z_{_{KdS_{1,2}}}\!\approx \left( 1+z_{_{Kerr_{1,2}}}\!\right) \left(
1+z_{\Lambda }\right) ,  \label{KdSvsKerr}
\end{equation}%
where $z_{\Lambda }=\sqrt{\Lambda /3}r_{d}$ is the contribution of the
cosmological constant in the redshift, and the factors $1+z_{_{Kerr_{1,2}}}$
have the following explicit forms 
\begin{equation}
1+z_{_{Kerr_{1}}}\!=\frac{\left( 1-2\tilde{M}\right) \pm \tilde{M}%
^{1/2}\left( \tilde{a}+\sqrt{\tilde{\Delta}_{Kerr}}\right) }{\left( 1-2%
\tilde{M}\right) \sqrt{1-3\tilde{M}\pm 2\,\tilde{a}\,\tilde{M}^{1/2}}},
\label{KerrRedFarDetector}
\end{equation}%
\begin{equation}
1+z_{_{Kerr_{2}}}\!=\frac{\left( 1-2\tilde{M}\right) \pm \tilde{M}%
^{1/2}\left( \tilde{a}-\sqrt{\tilde{\Delta}_{Kerr}}\right) }{\left( 1-2%
\tilde{M}\right) \sqrt{1-3\tilde{M}\pm 2\,\tilde{a}\,\tilde{M}^{1/2}}},
\label{KerrBlueFarDetector}
\end{equation}%
that are the frequency shifts in the standard Kerr spacetime found in \cite%
{PRDbhmn} with $\tilde{M}=M/r_{e}$, $\ \tilde{a}=a/r_{e}$, and $\tilde{\Delta%
}_{Kerr}=1+\tilde{a}^{2}-2\tilde{M}$.

The Hubble constant $H_{0}$ is related to the cosmological constant with 
\cite{Carroll}%
\begin{equation}
H_{0}=\sqrt{\frac{\Lambda }{3\Omega _{\Lambda }}},  \label{HLrelation}
\end{equation}%
where $\Omega _{\Lambda }$\ is the cosmological constant density parameter.
For the special case of $\Omega _{\Lambda }=1$ (the Universe filled with
dark energy, i.e. in the absence of matter), we recover the Hubble law from $%
z_{\Lambda }=\sqrt{\Lambda /3}\,r_{d}$ as%
\begin{equation}
z_{\Lambda }=H_{0}\,r_{d},
\end{equation}%
in which $z_{\Lambda }$\ represents the velocity of the host galaxy going
away from the detector and $r_{d}$ is the distance between the black hole
and the observer. By introducing the relation (\ref{HLrelation}) in Eq. (\ref%
{KdSvsKerr}), we can obtain the frequency shift in the KdS background in
terms of the Kerr black hole parameters and the Hubble constant as below%
\begin{equation}
1+z_{_{KdS_{1,2}}}\!\approx \left( 1+z_{_{Kerr_{1,2}}}\!\right) \left( 1+%
\sqrt{\Omega _{\Lambda }}\,H_{0}\,r_{d}\right) ,  \label{ZvsH0}
\end{equation}%
an expression that can be employed to obtain $H_{0}$\ as well as black hole
parameters. Therefore, we extracted the Hubble law by considering the
frequency shift of stars orbiting around Kerr black holes in asymptotically
dS spacetime detected by a far-away observer.

Note that the formula (\ref{ZvsH0})\ does not constitute a simple
multiplication of $1+z_{_{Kerr_{1,2}}}$\ and $1+\sqrt{\Omega _{\Lambda }}%
\,H_{0}\,r_{d}$ by hand, since here the second factor arose quite naturally
as a dominant term of the cosmological redshift from general relativistic
considerations, while the most general and complicated expressions are given
through the rhs of Eqs. (\ref{GeneralRed})-(\ref{GeneralBlue}). This formula
suggests that the photons emitted from massive geodesic particles revolving
KdS black hole contain information about mass $M $, spin $a$, and the
cosmological constant $\Lambda $\ when they arrive at the detector. This
information is encoded in the frequency shift of these photons, denoted by $%
z_{_{KdS_{1,2}}}$ in Eq. (\ref{ZvsH0}), and is a directly observable
quantity. Hence, the set of spacetime parameters $\left\{ M,a,\Lambda
\right\} $ can be estimated by performing a statistical fit \cite%
{ApJL,TXS,TenAGNs,FiveAGNs} and sometimes analytically expressed by solving
an inverse problem like in \cite{PRDbhmn}. In other words, the formula (\ref%
{ZvsH0}) allows extracting the properties of spacetime characterized by the
black hole mass $M$\ and spin $a$ as well as the cosmological constant $%
\Lambda $\ through measuring shifts in the frequency of photons $%
z_{_{KdS_{1,2}}}\!$.

\subsection{Schwarzschild black hole mass and the Hubble constant in terms
of redshift/blueshift}

\label{SchwSec}

Here, we obtain analytical formulas for the Schwarzschild black hole mass
and the Hubble constant in terms of frequency shifts of photons. Then, we
recover the Hubble law from the latter analytic expression.

For the static Schwarzschild black holes, the redshift formula (\ref{ZvsH0}%
)\ reduces to (for $\Omega _{\Lambda }=1$) 
\begin{equation}
1+z_{_{SdS_{1,2}}}\!\approx \left( 1+z_{_{Schw_{1,2}}}\!\right) \left(
1+H_{0}r_{d}\right) ,  \label{SdSvsH0}
\end{equation}%
where $z_{_{SdS_{1,2}}}$\ is the frequency shift of SdS black holes and $%
z_{_{Schw_{1,2}}}$\ is the frequency shift of the Schwarzschild black holes
that can be determined by taking the limit $\tilde{a}\rightarrow 0$\ in (\ref%
{KerrRedFarDetector})-(\ref{KerrBlueFarDetector}) as follows 
\begin{equation}
1+z_{_{Schw_{1,2}}}\!=\frac{1}{\sqrt{1-3\tilde{M}}}\left( 1\pm \sqrt{\frac{%
\tilde{M}}{1-2\tilde{M}}}\right) ,  \label{zSchw}
\end{equation}%
in which the upper (lower) sign refers to redshifted (blueshifted) particles 
\cite{SchwShift}. Now, one can use Eqs. (\ref{SdSvsH0}) and (\ref{zSchw}) to
get%
\begin{equation}
RB=\frac{\left( 1+H_{0}r_{d}\right) ^{2}}{1-2\tilde{M}},  \label{RBproduct}
\end{equation}%
\begin{equation}
\frac{R}{B}=\frac{1-\tilde{M}+2\sqrt{\tilde{M}\left( 1-2\tilde{M}\right) }}{%
1-3\tilde{M}},  \label{RBratio}
\end{equation}%
where $R=1+z_{_{SdS_{1}}}$\ and $B=1+z_{_{SdS_{2}}}$. As the next step, we
solve the first equation\ (\ref{RBproduct})\ to obtain the Schwarzschild
black hole mass as below 
\begin{equation}
\tilde{M}=\frac{RB-\left( 1+H_{0}r_{d}\right) ^{2}}{2RB},
\label{MassRelation}
\end{equation}%
in terms of the frequency shift and $H_{0}r_{d}$ product. It is worth
noticing that when the $H_{0}$ constant vanishes, we recover the mass
formula as a function of $R$ and $B$ obtained in \cite{PRDbhmn}. In order to
find the dependency of $H_{0}r_{d}$-term\ on the redshift, we replace (\ref%
{MassRelation}) in Eq. (\ref{RBratio}) and solve for $H_{0}$\ as 
\begin{equation}
H_{0}=\frac{1}{r_{d}}\left( -1+\frac{\left( R+B\right) \sqrt{RB}}{\sqrt{%
3R^{2}+3B^{2}-2RB}}\right) ,  \label{H0}
\end{equation}%
which gives the Hubble constant in terms of the frequency shift $R$ and $B$
of the massive geodesic particles on either side of the Schwarzschild black
hole as well as the detector distance to the black hole $r_{d}$. Therefore,
the $H_{0}\,r_{d}$ product appearing in (\ref{H0}) can be used to expressed
the mass relation (\ref{MassRelation}) in terms of the redshift and
blueshift only. Alternatively, from Eq. (\ref{RBratio}) it is
straightforward to obtain the following expression 
\begin{equation}
\tilde{M}=\frac{(R-B)^{2}}{3R^{2}+3B^{2}-2RB},  \label{M0}
\end{equation}%
for the mass parameter defined by purely observational quantities $R$ and $B$%
.

For the special case $\tilde{M}<<1$, we obtain $R\approx B$ (see Eq. (\ref%
{RBratio})). In this situation, the redshift $z_{_{SdS_{1}}}$ and blueshift $%
z_{_{SdS_{2}}}$ are almost equal $z_{_{SdS_{1}}}\approx z_{_{SdS_{2}}}\equiv
z$. Thus, we recover the Hubble law for this case from the analytic
expression (\ref{H0}) by taking the limit $R\rightarrow B$ as follows%
\begin{equation}
z=H_{0}\,r_{d}.
\end{equation}%
\ 

It is worthwhile to mention that finding a relation of the form (\ref%
{SdSvsH0}) [or Eq. (\ref{ZvsH0}) for the rotating case] has another
significant importance practically. For instance, in the case of accretion
disks circularly orbiting supermassive black holes in the center of AGNs
within the Hubble flow, the total redshift of emitted photons is given by%
\begin{equation}
1+z_{_{tot_{1,2}}}\!=\left( 1+z_{_{Schw_{1,2}}}\!\right) \left(
1+z_{rec}\right) ,
\end{equation}%
in which the recessional redshift of galaxies $z_{rec}$ is composed by \cite%
{Davis}%
\begin{equation}
1+z_{rec}=\left( 1+z_{Cosm}\right) \left( 1+z_{Boost}\right) ,
\end{equation}%
where $z_{Cosm}$\ is the cosmological redshift due to the accelerated
expansion of the Universe and $z_{Boost}$\ is a special relativistic
redshift due to the peculiar motion produced by the local gravity effects
(see \cite{ApJL,TXS,TenAGNs,FiveAGNs}\ when the geometry of the central
objects was described by the Schwarzschild\ line element). In this relation,
since $z_{Cosm}$\ and $z_{Boost}$\ do not depend on the metric, the
cosmological redshift and the peculiar redshift become degenerate and we can
just obtain $z_{rec}$, but not $z_{Cosm}$\ and $z_{Boost}$\ separately. On
the contrary, since the dependency of $z_{Cosm}$\ on the metric derived in (%
\ref{SdSvsH0}) as $z_{Cosm}=H_{0}r_{d}$ is explicit [or more completely in
Eqs. (\ref{GeneralRed})-(\ref{GeneralBlue})], this fact can help to break
the degeneracy between $z_{Cosm}$\ and $z_{Boost}$, allowing us to estimate
both of these frequency shifts separately.

As the final point, we would like to discuss how to measure the black hole
parameters by employing this formalism a little bit. In order to apply the
present method to real astrophysical systems, like the megamasers orbiting a
central black hole in AGNs, one can use the approximation $r_{e}\approx
\delta r_{d}$ such that $\delta $ is the aperture angle of the telescope
that is a measurable quantity. Thus, in the case of static Schwarzschild
black holes, the total redshift is a function of $%
z_{Schw_{1,2}}=z_{Schw_{1,2}}\left( M,r_{d}\right) $ and we can employ Eq. (%
\ref{M0}) to compute the mass-to-distance ratio $M/r_{d}$ in terms of the
observable quantities $z_{Schw_{1}}$ and $z_{Schw_{2}}$, as it was
accomplished for the central black hole of NGC 4258 in \cite{ApJL} and more
sixteen galaxies in \cite{TXS,TenAGNs,FiveAGNs}. Moreover, if the distance
to the central black hole of the galaxy $r_{d}$ is known from a different
astrophysical experiment, then we can determine the Schwarzschild black hole
mass $M$ alone (see \cite{TXS} for the TXS 2226-184 galaxy, for instance).

Alternatively, the total redshift is a function of $%
z_{SdS_{1,2}}=z_{SdS_{1,2}}\left( M,r_{d},\Lambda \right) $\ in the case of
SdS black holes, hence one can basically estimate $M$, $r_{d}$, and $\Lambda 
$ (or $H_{0}$) with the aid of a statistical fit for AGNs within the Hubble
flow. In this case, the Bayesian fitting method allows us to estimate $M$, $%
r_{d}$, and $\Lambda $\ separately (not the mass-to-distance ratio $M/r_{d}$%
) since the functional dependence of the total redshift on $r_{d}$ in Eqs. (%
\ref{SdSvsH0})-(\ref{zSchw}) is different in the first term and the second
term. With observational data detected from areas close enough to the black
hole at hand, in principle we can also try to estimate the rotation
parameter $a$ by employing Eqs. (\ref{KdSvsKerr})-(\ref{KerrBlueFarDetector}%
).

\section{Discussion and final remarks}

In this paper, we have taken into account the KdS solutions and analytically
obtained valid parameter space for having KdS black holes. Then, we have
expressed the frequency shift of photons emitted by massive geodesic
particles, stars for instance, that are circularly orbiting the KdS black
holes in terms of the parameters of spacetime, such as the black hole mass,
angular momentum, and cosmological constant. For this purpose, we have
considered the detectors to be in radial motion with respect to the
emitter-black hole system and employed a general relativistic formalism that
was briefly described through the text.

In addition, we have seen that the shift in frequency of photons increases
with an increase in the cosmological constant as well as the detector
distance to the emitter-black hole system that was compatible with the
repulsive nature of the cosmological constant. Hence, this observation led
us to extract the Hubble law from the original redshift formulas\ by taking
into account some physically motivated approximations.

Moreover, we have found analytic expressions for the Schwarzschild black
hole mass and the Hubble constant in terms of the observational frequency
shifts of massive particles orbiting circularly this static spherically
symmetric black hole. Interestingly, we have also shown that the Hubble law
arose naturally from the exact formula of the Hubble constant (\ref{H0}).
The concise and elegant formulas that we have found allow us to extract the
properties of spacetime characterized by the black hole mass and spin as
well as the cosmological constant through measuring shifts in the frequency
of photons.

Now, we finish our paper with a couple of suggestions for future work. It
would be interesting to employ and generalize this work in some other
directions. For instance, in this study, we were interested in emitters in
the range $r_{e}\in \lbrack r_{ISCO},r_{OSCO}]$ and far-away detectors
within $r_{d}\in (\bar{r},r_{c})$, describing the black hole systems in the
Hubble flow. However, this formalism can be generalized for circularly
orbiting (or static) detectors as well for possible local tests of the
accelerated expansion of the Universe. On the other hand, the formula (\ref%
{SdSvsH0})\ can be employed to estimate the Schwarzschild black hole mass $M$%
, the distance $r_{d}$ to the black hole, and the Hubble constant $H_{0}$\
(or the cosmological constant $\Lambda $) by using accretion discs
circularly orbiting supermassive black holes hosted at the core of AGNs with
the help of Bayesian fitting methods. Our primary estimations of $H_{0}$
based on the observational data of galaxies within the Hubble flow show that
this approach could be a powerful tool to obtain the Hubble constant
alongside the black hole parameters. This investigation is currently under
consideration.

Finally, we would like to stress that the $H_{0}$ expression, that we
obtained in (\ref{H0}) with the help of the KdS metric, represents a first
step towards a more realistic parametrization of $H_{0}$ in terms of
observable quantities that also considers the matter content of the
Universe, in consistency with the $\Lambda $-cold dark matter cosmological
standard model. We are currently studying this problem and hope to report on
it in the near future.

%%%%%%%%%%%%%%%%%%%%%%%%%%%%%%%%%%%%%%%%%%%%%%%%%%%%%%%%%%%%%%%%%%%%%%%%%%%%%%%%%

\section*{Acknowledgments}

All authors are grateful to CONACYT for support under Grant No.
CF-MG-2558591; M.M. also acknowledges CONACYT for providing financial
assistance through the postdoctoral Grant No. 31155. A.H.-A. and U.N. thank
SNI and PROMEP-SEP and were supported by Grants VIEP-BUAP No. 122 and No.
CIC-UMSNH, respectively. U.N. also acknowledges support under Grant No.
CF-140630.

%%%%%%%%%%%%%%%%%%%%%%%%%%%%%%%%%%%%%%%%%%%%%%%%%%%%%%%%%%%%%%%%%%%%%%%%%%%%%%%%%


\begin{thebibliography}{99}
\bibitem{GW} B. P. Abbott \textit{et al}. (LIGO Scientific and Virgo
Collaborations), Phys. Rev. Lett. 116, 061102 (2016).

\bibitem{EHTM87} K. Akiyama \textit{et al}. (Event Horizon Telescope
Collaboration), Astrophys. J. Lett. 875, L4 (2019).

\bibitem{EHTSgr} K. Akiyama \textit{et al}. (Event Horizon Telescope
Collaboration), Astrophys. J. Lett. 930, L17 (2022).

\bibitem{PRDhn} A. Herrera-Aguilar and U. Nucamendi, Phys. Rev. D 92, 045024
(2015).

\bibitem{PRDbhmn} P. Banerjee, A. Herrera-Aguilar, M. Momennia and U.
Nucamendi, Phys. Rev. D 105, 124037 (2022).

\bibitem{MyersPerry} M. Sharif and S. Iftikhar, Eur. Phys. J. C 76, 404
(2016).

\bibitem{KNdS} G. V. Kraniotis, Eur. Phys. J. C 81, 147 (2021).

\bibitem{PlebanskiDemianski} D. Ujjal, Chin. J. Phys. 70, 213 (2021).

\bibitem{RegularBH} R. Becerril, S. Valdez-Alvarado, U. Nucamendi, P.
Sheoran and J. M. Davila, Phys. Rev. D 103, 084054 (2021).

\bibitem{BosonStar} R. Becerril, S. Valdez-Alvarado and U. Nucamendi, Phys.
Rev. D 94, 124024 (2016).

\bibitem{MOG} P. Sheoran, A. Herrera-Aguilar and U. Nucamendi, Phys. Rev. D
97, 124049 (2018).

\bibitem{NED} L. A. Lopez and J. C. Olvera, Eur. Phys. J. Plus 136, 64
(2021).

\bibitem{SMF} L. A. Lopez and N. Breton, Astrophys. Space Sci. 366, 55
(2021).

\bibitem{FuZhang} Q. M. Fu and X. Zhang, Phys. Rev. D 107, 064019 (2023).

\bibitem{ApJL} U. Nucamendi, A. Herrera-Aguilar, R. Lizardo-Castro and O.
Lopez-Cruz, Astrophys. J. Lett. 917, L14 (2021).

\bibitem{TXS} A. Villalobos-Ramirez, O. Gallardo-Rivera, A. Herrera-Aguilar
and U. Nucamendi, Astron. Astrophys. 662, L9 (2022).

\bibitem{TenAGNs} D. Villaraos, A. Herrera-Aguilar, U. Nucamendi, G.
Gonzalez-Juarez and R. Lizardo-Castro, MNRAS 517, 4213 (2022).

\bibitem{FiveAGNs} A. Villalobos-Ramirez, A. Gonzalez-Juarez, M. Momennia
and A. Herrera-Aguilar, [arXiv:2211.06486].

\bibitem{DarkEnergy} S. M. Carroll, W. H. Press and E. L. Turner, Annu. Rev.
Astron. Astrophys. 30, 499 (1992).

\bibitem{Carroll} S. M. Carroll, Living Rev. Rel. 4, 1 (2001).

\bibitem{KdS} G. W. Gibbons and S. W. Hawking, Phys. Rev. D 15, 2738 (1977).

\bibitem{KdSECO} Z. Stuchlik and P. Slany, Phys. Rev. D 69, 064001 (2004).

\bibitem{Hackmann} E. Hackmann, C. Lammerzahl, V. Kagramanova and J. Kunz,
Phys. Rev. D 81, 044020 (2010).

\bibitem{Kraniotis} G. V. Kraniotis, Class. Quantum Grav. 28, 085021 (2011).

\bibitem{MCPI} M. J. Reid, J. A. Braatz, J. J. Condon, L. J. Greenhill, C.
Henkel and K. Y. Lo, Astrophys. J. 695, 287 (2009).

\bibitem{MCPIII} C. Y. Kuo, J. A. Braatz, J. J. Condon, C. M. V.
Impellizzeri, K. Y. Lo, I. Zaw, M. Schenker, C. Henkel, M. J. Reid and J. E.
Greene, Astrophys. J. 727, 20 (2011).

\bibitem{MCPV} C. Y. Kuo, J. A. Braatz, M. J. Reid, K. Y. Lo, J. J. Condon,
C. M. V. Impellizzeri and C. Henkel, Astrophys. J. 767, 155 (2013).

\bibitem{MCPVI} C. Y. Kuo, J. A. Braatz, K. Y. Lo, M. J. Reid, S. H. Suyu,
D. W. Pesce, J. J. Condon, C. Henkel and C. M. V. Impellizzeri, Astrophys.
J. 800, 26 (2015).

\bibitem{MCPVIII} F. Gao, J. A. Braatz, M. J. Reid, K. Y. Lo, J. J. Condon,
C. Henkel, C. Y. Kuo, C. M. V. Impellizzeri, D. W. Pesce and W. Zhao,
Astrophys. J. 817, 128 (2016).

\bibitem{MCPIX} F. Gao, J. A. Braatz, M. J. Reid, J. J. Condon, J. E.
Greene, C. Henkel, C. M. V. Impellizzeri, K. Y. Lo, C. Y. Kuo and D. W.
Pesce, Astrophys. J. 834, 52 (2017).

\bibitem{MCPXI} D. W. Pesce, J. A. Braatz, M. J. Reid, J. J. Condon, F. Gao,
C. Henkel, C. Y. Kuo, K. Y. Lo and W. Zhao, Astrophys. J. 890, 118 (2020).

\bibitem{MCPXII} C. Y. Kuo, J. A. Braatz, C. M. V. Impellizzeri, F. Gao, D.
Pesce, M. J. Reid, J. Condon, F. Kamali, C. Henkel and J. E. Greene, MNRAS
498, 1609 (2020).

\bibitem{MCPXIII} D. W. Pesce, J. A. Braatz, M. J. Reid, A. G. Riess, D.
Scolnic, J. J. Condon, F. Gao, C. Henkel, C. M. V. Impellizzeri, C. Y. Kuo
and K. Y. Lo, Astrophys. J. Lett. 891, L1 (2020).

\bibitem{CarterCons} B. Carter, Phys. Rev. 174, 1559 (1968).

\bibitem{BardeenPetterson} J. M. Bardeen and J. A. Petterson, Astrophys. J.
Lett. 195, L65 (1975).

\bibitem{Darling} J. Darling, Astrophys. J. 100, 837 (2017).

\bibitem{SchwShift} It does not matter whether the particle is co-rotating
or counter-rotating due to the absence of the dragging effect produced by
the rotation nature of spacetime. For more details see Fig. 3 of \cite%
{PRDbhmn} and the related discussion.

\bibitem{Davis} T. M. Davis and M. I. Scrimgeour, MNRAS 442, 1117 (2014).
\end{thebibliography}
\end{document}